\documentclass[aps,preprint,letterpaper]{revtex4}
\usepackage{bm}
\usepackage{amsmath}
\usepackage{amsthm}
\usepackage{dsfont}
\usepackage{verbatim}
\usepackage{bm,braket}
\usepackage{color}
\usepackage{float}
\usepackage{graphicx}
\usepackage{multirow}
\usepackage{array}
\usepackage{amssymb}
\usepackage{sidecap}
\usepackage{enumitem}
\usepackage[toc,page]{appendix}
\usepackage{epstopdf}
\usepackage[caption=false]{subfig}
\usepackage{mathtools}

\usepackage{sidecap}
\usepackage{enumitem}
\usepackage{mathtools}
\usepackage{epstopdf}
\usepackage{subfig}
\usepackage{sidecap}
\usepackage[T1]{fontenc}
\usepackage[utf8]{inputenc}
\usepackage{tikz}
\usepackage[version=4]{mhchem}

\usepackage{microtype}

\usepackage{bm}
\usepackage{amsmath}
\usepackage{verbatim}
\usepackage{bm,braket}
\usepackage{color}
\usepackage{float}
\usepackage{graphicx}
\usepackage{multirow}
\usepackage{array}
\usepackage{amssymb}
\usepackage{sidecap}
\usepackage{enumitem}
\usepackage[toc,page]{appendix}
\usepackage{epstopdf}
\usepackage[caption=false]{subfig}
\usepackage{sidecap}
\usepackage{enumitem}
\usepackage[toc,page]{appendix}
\usepackage{epstopdf}
\usepackage{subfig}
\usepackage{sidecap}
\usepackage[utf8]{inputenc}
\usepackage[compact]{titlesec}
\usetikzlibrary{arrows,shapes,decorations.pathmorphing}
\usetikzlibrary{decorations.pathmorphing,decorations.pathreplacing,decorations.shapes}
\usepackage[makeroom]{cancel}

\makeatletter

\definecolor{myBlue}{rgb}{0.0430,0.5156,0.7773}
\definecolor{light-gray}{gray}{0.95}

\newcommand{\newket}[1]{|{#1}\rangle \rangle}

\newcommand{\lket}[1]{\left|{#1}\right)}

\newcommand{\lbraket}[1]{\left({#1}\right)}

\newcommand*\bigcdot{\mathpalette\bigcdot@{.8}}
\newcommand*\bigcdot@[2]{\mathbin{\vcenter{\hbox{\scalebox{#2}{$\m@th#1\bullet$}}}}}
\makeatother

\usepackage[breaklinks=true,linkcolor=blue]{hyperref}
\usepackage{xcolor}
\hypersetup{
	colorlinks,
	linkcolor={black},
	citecolor={blue!50!black},
	urlcolor={black}
}

\usepackage{natbib}
\usepackage[english]{babel}

\begin{document}

	\frenchspacing
	\abovedisplayskip=8pt
	\abovedisplayshortskip=8pt
	\belowdisplayskip=8pt
	\belowdisplayshortskip=8pt
	\arraycolsep=100pt
	\setlength{\parskip}{5pt}
	\setlength{\parindent}{16pt}
	
	\title{An efficient approach to  the quantum dynamics and rates of processes induced by natural incoherent light}
	
	\author{Simon Axelrod}
	\affiliation{Chemical Physics Theory Group, Department of Chemistry, and Center for Quantum Information and Quantum Control, University of Toronto, Toronto, Ontario M5S 3H6, Canada}
	\author{Paul Brumer}
	\affiliation{Chemical Physics Theory Group, Department of Chemistry, and Center for Quantum Information and Quantum Control, University of Toronto, Toronto, Ontario M5S 3H6, Canada}
	\date{\today}
	
	\begin{abstract}
		In many important cases, the rate of excitation of a system embedded in an environment is significantly smaller than the internal system relaxation rates. An important example is that of light-induced processes under natural conditions, in which the system is excited by weak, incoherent (e.g., solar) radiation. Simulating the dynamics on the time scale of the excitation source can thus be computationally intractable. Here we describe a method for obtaining the dynamics of quantum systems without directly solving the master equation. We present an algorithm for the numerical implementation of this method, and, as an example, use it to reconstruct the internal conversion dynamics of pyrazine excited by sunlight. Significantly, this approach also allows us to assess the role of quantum coherence on biological time scales, which is a topic of ongoing interest.
	\end{abstract}
	
	\maketitle

 \section{Introduction}
 Natural light-induced processes such as photosynthesis \cite{photo1,photo2}, vision \cite{exp3,exp4,hoki,fs1}, and photocell function \cite{fano1,fano2,fano4}, as well as noise-induced dynamics \cite{noise1,noise2}, are characterized by continuous, weak excitation. For example, the excitation rate associated with solar illumination and a typical transition dipole moment of 10 D is on the order of an inverse microsecond \cite{timur2}. The time scale grows larger as the intensity of radiation is reduced \cite{hoki}. By contrast, the internal dynamics associated with light-harvesting complexes \cite{photo1,photo2} or light-sensing molecules such as the retinal chromophore \cite{fs1,fs3,miller_speed_limit,exp3}, as induced in pulsed laser experiments, occur on a femtosecond time scale. Thus simulating the full dynamics of such natural processes in open systems (i.e. systems coupled to an environment) could require solving a master equation with $m$ time steps of size $\Delta t$, such that $\Delta t  \sim \mathcal{O}( \mathrm{fs})$, and $m \Delta t  \sim \mathcal{O}(\mu \mathrm{s})$. That is, the required number of time-steps would exceed $10^9$. Even if the system equilibrates on a shorter time scale determined by the relevant relaxation processes, such as spontaneous emission, non-radiative population decay \cite{balzer}, or energy transfer to an adjacent chromophore \cite{jesenko}, a full simulation could also be prohibitive. For example, if the time scale of spontaneous emission is on the order of a nanosecond (as is the case for a dipole moment of 10 D and an optical transition), then the number of time-steps would exceed $10^6$. 
 
 However, if one is only interested in reaction rates, then a full simulation of the process may be unnecessary. Since the rate of excitation is expected to be small compared to internal relaxation rates, the dynamics are expected to be approximately exponential after a certain transient turn-on time \cite{2001,kapral,passage_review_2}. That is, in a generic reaction of the form $R \rightleftharpoons P$, with $R$ being the reactant and $P$ being the product, the quantity $R(t)- R_{\mathrm{eq}}$ is expected to decay as $e^{- k  t}$ \cite{2001}. Here, $R(t)$ is the probability that the system is in product form, $R_{\mathrm{eq}} = R(t \to \infty)$ is the equilibrium value of $R$, and $k$ is the rate constant. In this case, it is only necessary to simulate past the transient time to the time $t^{*}$, where $t_{\mathrm{mic}} \ll t^* \ll t_{\mathrm{chem}}$, with $t_{\mathrm{mic}}$ being the fast microscopic time scale, and $t_{\mathrm{chem}}$ being the slow chemical time scale \cite{2001,kapral}. By the time $t^*$, the reaction is expected to proceed in the aforementioned exponential fashion, and the rate of the process may be obtained directly from simulation. The time $t^*$ is much less than the reaction time scale, but much greater than the internal relaxation time scale, and thus simulation requires far fewer time-steps. Indeed, simulation to time $t^*$ may be possible for systems with small Hilbert spaces. It may also be possible for systems with large Hilbert spaces if the associated master equation has computationally favorable properties. For example, the canonical example of a system that is weakly coupled to a thermal bath and described by a second-order Born-Markov master equation may be simplifiable by the ``secular approximation'' \cite{breuer}. In this case, diagonal elements of the density matrix in the energy eigenbasis couple only to one another, while the evolution of off-diagonal elements is known analytically \cite{hoki}. Thus for a Hilbert space of dimension $d$, the differential equation that is propagated has only $d$ unknowns, rather than $d^2$. 
 
 However, systems with large Hilbert spaces subject to optical excitation may not be accurately described by the secular approximation \cite{psbr,timur2,vutha}. Moreover, the secular approximation does not allow for the generation of coherences, i.e. off-diagonal elements of the density operator in the energy eigenbasis, if there are none present initially. Given the ongoing interest and debate regarding the role of quantum coherences in biological processes \cite{coherence_photosynthesis1,coherence_photosynthesis2,coherence_photosynthesis3,exp3,exp4}, it is often crucial to go beyond the secular approximation. However, when the secular approximation is not invoked, a simulation to time $t^*$ may be impractical.
 In such cases it is desirable to obtain the reaction rate without directly solving the master equation. Further, if we also wish to probe the dynamics at long times when the reaction may not be precisely exponential, then solving the full master equation would certainly be prohibitive.

 Motivated by these considerations, we develop a method below for reconstructing the time dependence of the reaction dynamics and the associated rates without direct solution of the quantum master equation. In this way we can obtain reaction rates, analyze long-time non-exponential behavior, and consider the role of coherences in the reaction dynamics. Our approach is reminiscent of the first passage time technique in classical dynamics \cite{original_passage}, but is completely quantum mechanical in nature.
 In section \ref{sec:theory} we present our method and connect the results to chemical rate law phenomenology. In section \ref{sec:example_systems} this method is applied to two analytically soluble model systems. Section \ref{sec:numerics} provides a detailed computational algorithm that can be used to apply our technique. Sections \ref{sec:example_systems} and \ref{sec:numerics} are rather technical and may be skipped by readers interested in an overview of the method only. The algorithm is used to reconstruct the dynamics of internal conversion in model pyrazine in section \ref{sec:pyrazine}. Section \ref{sec:conclusion} summarizes the work.
\section{Theory}
\label{sec:theory}
\subsection{Reconstruction of the progress variable}
\label{subsection:progress_moments}
\subsubsection{Time-independent Liouville superoperator}
To follow the dynamics, consider the operator $\hat{\chi} =\hat{O}- \mathrm{Tr}(\hat{\rho}_s \hat{O}) \hat{\mathds{1}}$, where $\hat{\rho}_s = \hat{\rho}(t\to \infty)$ is the steady state density operator in the Schr\"odinger picture, $\hat{\mathds{1}}$ is the identity operator, and $\hat{O}$ is an operator that monitors a physical process. The steady state density operator satisfies $\partial_t \hat{\rho}_s \equiv \hat{\mathcal{L}}\hat{\rho}_s = \hat{0}$, where $\hat{\mathcal{L}}$ is the Liouville superoperator, assumed time-independent.
As an example of a progress operator, consider a reaction of the form $R\rightleftharpoons P$, where $R$ denotes the reactant, $P$ denotes the product, and $\hat{R}$ and $\hat{P}$ are the associated projection operators onto reactant and product species, respectively. In this case $\hat{\chi}$ can be given by either $\hat{R} - \mathrm{Tr}[\hat{\rho}_s \hat{R}]\hat{\mathds{1}}$ or $\hat{P} - \mathrm{Tr}[\hat{\rho}_s \hat{P}]\hat{\mathds{1}}$. 

Interest is in reconstructing the dynamics of the expectation value of the progress operator, $\braket{\hat{\chi}(t)}  = \mathrm{Tr}[\hat{\rho}(t) \hat{\chi}]$, which we term the progress variable. In order to gain information about the reaction dynamics, consider the following quantity, termed the $n$th progress moment:
\begin{align}
I_{n} \equiv \int_{0}^{\infty} dt \  t^n \braket{\hat{\chi}(t)}, \label{eq:integral}
\end{align}
where $n$ is a non-negative integer. The time $t=0$ defines the beginning of the dynamics; for example, in light-induced processes, it defines the time at which the molecule and the solar radiation first interact. To evaluate this integral for arbitrary $\hat{\chi}$, we introduce the integral
\begin{align}
\delta \hat{\rho}_n \equiv \int_{0}^{\infty} dt \ t^n [\hat{\rho}(t) -\hat{\rho}_s], \label{eq:delta_rho_n}
\end{align}
such that $I_n =  \mathrm{Tr}[\hat{O}(\delta \hat{\rho}_n)]$. Applying $\hat{\mathcal{L}}$ to both sides of Eq. (\ref{eq:delta_rho_n}), using the master equation $\hat{\mathcal{L}} \hat{\rho}(t) = \partial_t \hat{\rho}(t) $, and integrating by parts yield
\begin{align}
\hat{\mathcal{L}}   [\delta \hat{\rho}_n] =   \int_{0}^{\infty} dt \ t^n \ \partial_t(\hat{\rho}(t) - \hat{\rho}_s) \equiv -\hat{c}_n =  -\begin{cases}
\hat{\rho}_0 - \hat{\rho}_s, & \text{if } n= 0\\
n \cdot \delta \hat{\rho}_{n-1}  &\text{if } n \neq 0,
\end{cases} \label{eq:delta_rho_1}
\end{align}
where $\hat{\rho}_0=\hat{\rho}(0)$ is the initial density operator.
To obtain this result, note that when integrating by parts, the lower boundary terms vanish at $t=0$ because $t^{m} \vert_{t=0} = 0$ for $m >0 $, and the upper boundary terms vanish since $\hat{\rho}(t\to \infty) - \hat{\rho}_s= \hat{0}$ \cite{vanishing_footnote}. These quantum recursive relations [Eq. (\ref{eq:delta_rho_1})] are reminiscent of those used to calculate $n$th passage times in classical barrier crossing problems \cite{passage_review_2}. 

Since there exists a non-trivial solution $\hat{\rho}_s$ to the equation $\hat{\mathcal{L}} \hat{\rho}_s = \hat{0}$, the superoperator $\hat{\mathcal{L}}$ is singular and, therefore, $\delta \hat{\rho}_n$ cannot be calculated by inverting $\hat{\mathcal{L}}$ in Eq. (\ref{eq:delta_rho_1}). However, since $\mathrm{Tr}[\hat{\rho}(t)]=1$ at all times, $\mathrm{Tr}[\delta \hat{\rho}_n] = 0$. This constraint can be incorporated into Eq. (\ref{eq:delta_rho_1}) by adding $w \hat{\mathcal{T}} [\delta \hat{\rho}_n]$ to the left-hand side, where $w$ is an arbitrary constant, $\hat{\mathcal{T}}$ is a superoperator that acts on $\delta \hat{\rho}_n$ through $\hat{\mathcal{T}} [ \delta \hat{\rho}_n] = \sum_{ijkl} \mathcal{T}_{ijkl} \bra{k} \delta \hat{\rho}_n \ket{l} \ket{i} \bra{j}$, and $\ket{\bigcdot}$ denotes a basis vector. The superoperator $\hat{\mathcal{T}}$ has components $\mathcal{T}_{ijkl} = \delta_{kl} \delta_{ij} \delta_{i1}$ \cite{steady_methods}, where $\ket{1}$ is an arbitrary basis vector and $\delta_{ij}$ is the Kronecker delta. Therefore, the effect of $w \hat{\mathcal{T}}$ is given by $w \sum_{ijkl} \delta_{kl} \delta_{ij} \delta_{i1} \braket{k \vert \delta \hat{\rho}_n \vert l} \ket{i} \bra{j} = w \mathrm{Tr}[\delta \hat{\rho}_n] \ket{1} \bra{1}$ for any constant $w$. 

Consider the addition of $w \hat{\mathcal{T}}[\delta \hat{\rho}_n]$ to the left-hand side of Eq. (\ref{eq:delta_rho_1}):
\begin{align}
\hat{\mathcal{L}}   [\delta \hat{\rho}_n] + w \hat{\mathcal{T}} [\delta \hat{\rho}_n] = -\hat{c}_n. \label{eq:wT_change}
\end{align}
Taking the trace of each side of Eq. (\ref{eq:wT_change}), using the fact that $\mathrm{Tr}[\hat{\rho}_0] - \mathrm{Tr}[\hat{\rho}_s]=0$, and invoking the identity $\mathrm{Tr}(\hat{\mathcal{L}}[\hat{O}]) = 0$ for any operator $\hat{O}$ (proved in Appendix \ref{app:non_markov}) yield $w \mathrm{Tr}[\delta \hat{\rho}_0] =0$. To obtain this result, we have used the fact that $\mathrm{Tr}(\ket{1}\bra{1}) = 1$ for any basis vector $\ket{1}$. Thus for any non-zero $w$ and any basis vector $\ket{1}$, the addition of addition of $w \hat{\mathcal{T}}$ implies that $\mathrm{Tr}[\delta \hat{\rho}_0] = 0$. The same result holds for $n\neq 0$, since $\mathrm{Tr}[\delta \hat{\rho}_{n-1}]=0$ by construction. Substituting this result back into Eq. (\ref{eq:wT_change}) recovers Eq. (\ref{eq:delta_rho_1}). Hence Eq. (\ref{eq:wT_change}) implies both Eq. (\ref{eq:delta_rho_1}) and $\mathrm{Tr}[\delta \hat{\rho}_n] = 0$. Moreover, the superoperator $\hat{\mathcal{L}} + w \hat{\mathcal{T}}$ is invertible (proved in Appendix \ref{app:non_markov}), and so $\delta \hat{\rho}_n$ is solved as $\delta \hat{\rho}_n = -[\hat{\mathcal{L} } + w \hat{\mathcal{T}}]^{-1} \hat{c}_n$  \cite{T_footnote}.

Once the progress moments $I_n$ are obtained, they can be used to reconstruct the progress variable $\braket{\hat{\chi}(t)}$. To see this, note that any function $g(t)$ defined for $t \geq 0$ satisfying $g(t\to \infty)=0$ can be written in a basis of decaying exponential functions \cite{amr},
\begin{align}
	g(t) = \int_{0}^{\infty} dk \ f(k) \ e^{-kt}, \label{eq:exp_basis}
\end{align}
subject to the initial condition $g(0) = \int dk \ f(k)$. The function $f(k)$ weights each exponential that decays at a rate $k$.
It has recently been shown that a wide variety of functions may be accurately represented with a limited number of exponential basis functions \cite{exp_basis1,exp_basis2}. Evidently, the number of required basis functions is especially small when $g(t)$ is approximately exponential. The progress moments $I_n$ are thus used to reconstruct $\braket{\hat{\chi}(t)}$ by projecting it onto a basis of exponentials. Setting $g(t) = \braket{\hat{\chi}(t)}$ and integrating Eq. (\ref{eq:exp_basis}) yield
\begin{align}
	I_n = \ n! \int_{0}^{\infty} dk \ f(k) \ k^{-(n+1)}.
\end{align}
Discretizing the set of basis functions as $f(k) dk \to \left\{  f_{m} \right\}$ and $k \to \left\{ k_m \right\}$ yields the expression
\begin{align}
	\braket{\hat{\chi}(t)} = \sum_m f_m e^{-k_m t}, \label{eq:expansion}
\end{align}
with the $k_m$ and $f_m$ defined through
\begin{align}
	&\sum_{m=0}^{m_{\mathrm{max}}} f_m \left(k_m \right)^{-n}=
	\begin{cases}
		\braket{\hat{\chi}(0)}, & \text{if } n= 0\\
		I_{n-1}/(n-1)!& \text{if } n \neq 0.
	\end{cases}
	\label{eq:basis_solve}
\end{align}
Here we have re-indexed $n$ as $n \in [0,n_{\mathrm{max}}]$, where $n_{\mathrm{max}}$ is the maximum number of computed progress moments, and $m_{\mathrm{max}}+1$ is the number of basis functions.

Hence, the goal of reconstructing the system dynamics has been reduced to three problems: first, solve $\hat{\mathcal{L}}\hat{\rho}_s = \hat{0}$ for the stationary state $\hat{\rho}_s$; second, solve Eq. (\ref{eq:delta_rho_1}) for $\delta \hat{\rho}_n$; and third, solve Eq. (\ref{eq:basis_solve}) for $f_m$ and $k_m$. A numerical method for dealing with several of these steps is given in section \ref{sec:numerics}.  An alternative to using progress moments is to use Laplace transforms of the progress variable, $\int_{0}^{\infty} dt \ \braket{\hat{\chi}(t)} e^{-k_n t}$, and to calculate the associated weights $f_m$ from the transformations. This approach is discussed in Appendix \ref{app:laplace}.  

The method based on progress moments is quite accurate when the progress variable decays on a single overall time scale, but less accurate when the progress variable decays on two different gross time scales. For example, it is quite accurate when the progress variable decays on a nanosecond time scale. However, when the progress variable is characterized by two decays, one on a nanosecond time scale and another on a millisecond time scale, the accuracy of the reconstructed dynamics on the nanosecond scale suffers. In this case, the nanosecond decay rates obtained from the progress variables can be made more accurate through the addition of a single Laplace transform, as discussed in Appendix \ref{app:laplace}.
\subsubsection{Generalizations}
The discussion above assumes that the Liouville superoperator is time-independent, an assumption that is valid for Markovian systems. In Appendix \ref{app:non_markov}, we show how these results can be generalized to a specific class of non-Markovian systems, where the density operator evolves according to $\partial_t \hat{\rho}(t) = \int_{0}^{t} d\tau \ \hat{\mathcal{K}}(t-\tau) \hat{\rho}(\tau)$ in the absence of initial system-bath correlations \cite{breuer,time_convolution_1,time_convolution_2}, with $\hat{\mathcal{K}}(t)$ being the memory kernel. The operator $\delta \hat{\rho}_n$ is then obtained as
\begin{align}
\delta \hat{\rho}_n = - [ \hat{\mathcal{L}}_0 + w \hat{\mathcal{T}} ]^{-1 } \begin{dcases}
\hat{\rho}_0 - \hat{\rho}_s  + \hat{T}_0,  \vphantom{\frac{0}{0}} & \text{if } n= 0 \\
n \cdot \delta \hat{\rho}_{n-1} + \hat{T}_n +  \sum_{k=0}^{n-1} \binom{n}{k} \hat{\mathcal{L}}_{n-k} [\delta \hat{\rho}_k],  &\text{if } n \neq 0,
\end{dcases}  \label{eq:extension}
\end{align}
where $\hat{\mathcal{L}}_n = \int_{0}^{\infty} dt \ t^n \ \hat{\mathcal{K}}(t)$, $\hat{T}_n = \int_{0}^{\infty} dt \ t^n ( \int_{0}^{t} d\tau \ \hat{\mathcal{K}}(\tau) \hat{\rho}_s )$ and $\binom{n}{k} = n!/[k! (n-k)!]$ is the binomial coefficient. In Appendix \ref{app:heom}, we provide a generalization of Eq. (\ref{eq:delta_rho_1}) to the non-Markovian, non-perturbative hierarchical equations of motion (HEOM), which do not involve an explicit memory kernel. However, in the main text we focus on dynamics that can be approximated as Markovian, which is a reasonable restriction if the progress variable evolves on a time scale that is considerably longer than the bath relaxation time (see Appendix \ref{app:laplace}).  Non-Markovian effects will be explored in future work.
\subsection{Chemical rate law phenomenology}
To provide insight into these expressions, consider the specific example of a chemical reaction whose products and reactants are separated by a large potential barrier. For a simple reaction of the form $R \rightleftharpoons P$, the phenomenological chemical rate law is \cite{2001}
\begin{align}
	\frac{d}{dt} \braket{\hat{R}(t)}  = - k_f \braket{\hat{R}(t)}  + k_r (1-\braket{\hat{R}(t)} ), \label{eq:rate_law}
\end{align}
where $k_f$ is the forward reaction rate and $k_r$ is the reverse rate. An appropriate choice for the progress variable is $\braket{\hat{\chi}(t) } = \braket{\hat{R}(t)} - \mathrm{Tr}[\hat{\rho}_s \hat{R}]$. If the time taken to climb the potential barrier between reactants and products is much longer than the time taken for internal relaxation on either side of the barrier, the chemical reaction is termed a rare event \cite{2001}. When such a separation of time scales holds, the reaction is expected to follow the exponential dynamics given by Eq. (\ref{eq:rate_law}) \cite{2001}. In this case, $\partial_t \braket{\hat{\chi}(t)} \approx - k \braket{\hat{\chi}(t)}$ at times $t>t^*$ \cite{kapral}, where the rate $k = k_{f} + k_{r}$ is a sum of forward and reverse reaction rates. The forward and reverse rates are related to $k$ as $k_f = kK/(1+K)$ and $k_r = k/(1+K)$, where $K = \mathrm{Tr}[\hat{\rho}_s \hat{P}]/\mathrm{Tr}[\hat{\rho}_s \hat{R}]$ is the equilibrium constant. 

To obtain $k$ through the method in section \ref{subsection:progress_moments}, consider that for such an exponential process, the zeroth progress moment $I_0$ would be given by $I_0 = \braket{\hat{\chi}(0)} k^{-1}$. This follows from the assumption that non-exponential dynamics for $t<t^*$ contribute negligibly to the integral in Eq. (\ref{eq:integral}). For such an exponential process, then, $ k \approx k^{(0)}$, where
\begin{align}
k^{(0)} \equiv [I_0 / \braket{\hat{\chi}(0)}]^{-1}. \label{eq:k_0}
\end{align}
The lowest order estimate of $k$ can therefore be obtained by solving for the stationary state $\hat{\rho}_s$, solving Eq. (\ref{eq:delta_rho_1}) for $\delta \hat{\rho}_0$, and calculating $k \approx k^{(0)}$ using $I_0 = \mathrm{Tr}[ \hat{O}(\delta \hat{\rho}_0)]$. The lowest order estimate of the reaction rate is thus obtained without solving a quantum master equation \cite{flux_footnote}. 

Note that while the zeroth moment $I_0$ is unaffected by early non-exponential dynamics, it may be affected by long time non-exponential behavior. A more accurate expression for the chemical reaction rate at intermediate times may then be obtained by solving Eqs. (\ref{eq:delta_rho_1}) and (\ref{eq:basis_solve}), and expressing the progress variable in the exponential basis of Eq. (\ref{eq:expansion}). The effect of long time non-exponential dynamics can be significant, in which case progress moments higher than the zeroth moment will be required to evaluate the reaction rate.
\section{Sample systems}
\label{sec:example_systems}
We discuss three sample systems as an example of this rate formalism. In this section, we consider two different model three-level systems and demonstrate analytical agreement. Numerical implementation of this approach and an example of the internal conversion of model pyrazine are described in Sections \ref{sec:numerics} and \ref{sec:pyrazine}.
\begin{figure}[t]
\includegraphics[width=0.9\columnwidth]{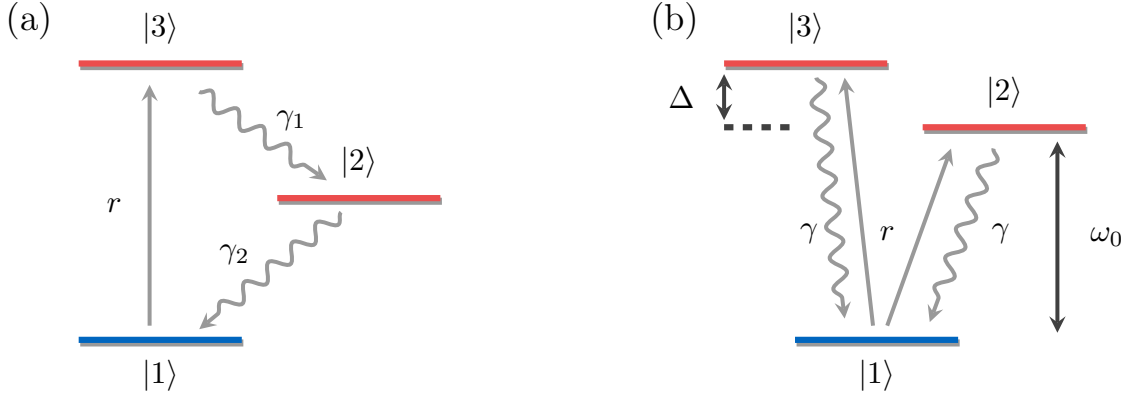}

	
	\caption{
		Energy diagram of model three-level systems. Excitation with rate $r$ is shown with straight gray lines, and relaxation with rates $\gamma_i$ are shown with wavy lines. The excited states are visualized in red and the ground state in blue. (a) A three-level system with coupled excited states, and (b) a $V$-system with uncoupled excited states separated by an energy splitting $\Delta \ll \omega_0$, where $\omega_0$ is the excitation frequency.}
	\label{fig:systems_schematic}
	
\end{figure}
\subsection{Three-level system}
\label{subsection:example_systems_chemical}
Consider first a three-level system with  incoherent pumping from ground state $\ket{1}$ to excited state $\ket{3}$ at rate $r$, with a decay from $\ket{3}$ to $\ket{2}$ at rate $\gamma_{1}$ and a decay from $\ket{2}$ to $\ket{1}$ at rate $\gamma_2$, as shown in Fig. \ref{fig:systems_schematic}(a). The levels are taken to be sufficiently separated in energy that the secular approximation is valid \cite{vutha}. Incoherent pumping and decay can then be represented by a set of Lindblad operators that decouple populations from coherences. The Lindblad operators are $ \hat{L}_k =   \sqrt{r} \ket{3}\bra{1} , \sqrt{\gamma_1} \ket{2}\bra{3}, \sqrt{\gamma_2} \ket{1}\bra{2} $, and the Lindblad equation is \cite{breuer}
\begin{align}
\frac{d}{dt} \hat{\rho} = -\frac{i}{\hbar} [\hat{H}_{\mathrm{S}}, \hat{\rho}]  + \sum_k \hat{L}_k \hat{\rho} \hat{L}_k^{\dagger} - \frac{1}{2} \{\hat{L}_k^{\dagger}  \hat{L}_k ,\hat{\rho} \},
\end{align} with $\hat{H}_{\mathrm{S}}$ being the system Hamiltonian and $\{ \ , \ \}$ being the anti-commutator.
To keep the analysis general, we write $\hat{\chi}$ in terms of its matrix elements, through $\hat{\chi} = \sum_{ij} O_{ij} \ket{i} \bra{j} -\mathrm{Tr}[\hat{\rho}_s \hat{O}] \hat{\mathds{1}}$. Defining $t_{ij} = O_{ii} - O_{jj}$ for $i,j = \ 1, \ 2, \ 3$, and taking $\hat{\rho}_0 = \ket{1}\bra{1}$, the zeroth progress moment is obtained as
\begin{align}
 [k^{(0)}]^{-1 } \equiv I_0 / \braket{\hat{\chi}(0)} =  \frac{\gamma_2^2 t_{13} + \gamma_1^2 t_{12} + \gamma_1 \gamma_2 t_{12} +\gamma_1 r t_{32}}{\left( \gamma_2 t_{13} + \gamma_1 t_{12}  \right) \left( \gamma_2 r + \gamma_1 r + \gamma_1 \gamma_2 \right) }, \label{eq:rate_3level}
\end{align}
where we have solved $\hat{\mathcal{L}}\hat{\rho}_s = \hat{0}$ and $\mathrm{Tr}[\hat{\rho}_s]=1$ to obtain $\hat{\rho}_s = (\gamma_1 \gamma_2 \ket{1} \bra{1} + \gamma_1 r \ket{2} \bra{2} + \gamma_2 r \ket{3} \bra{3})/(\gamma_1 \gamma_2 + \gamma_1 r + \gamma_2 r)$ and used the solution to solve Eq. (\ref{eq:delta_rho_1}). Equation (\ref{eq:rate_3level}) can be shown to be the solution obtained from solving the master equation. 

For example, if states $\ket{1}$ and $\ket{3}$ represent the reactant, and state $\ket{2}$ represents the product, then the operator of interest would be $\hat{O} = \ket{2} \bra{2}$. In this case, $t_{13} = 0$, while $ t_{12} = t_{32} =-1$. Furthermore, if $\gamma_2 = 0$, the reaction proceeds to completion, with $\hat{\rho}_s = \ket{2} \bra{2}$ as the steady state. Then in the limit that pumping is the rate-determining step ($r \ll \gamma_1$), the estimate $k \approx k^{(0)}$ yields $k = r$. Using the relation $k_f = kK/(1+K)$, with $k_f$ being the forward reaction rate and $K$ being the equilibrium constant, yields $k_f = r$ as expected. However, if $\gamma_2$ is non-zero, then the reaction does not proceed to completion. Then on the opposite extreme, where $\gamma_1, \gamma_2 \gg r$, the lowest order estimate $k^{(0)} \approx k$ gives
\begin{align}
k = \frac{\gamma_2 }{1 + \gamma_2/\gamma_1} .
\end{align}
In this limiting case, the equilibrium constant is given approximately by $K=r/\gamma_2$. In this case, the pumping rate $r$ competes with the rate $\gamma_2$, which removes population from the product state, and the relative strength of $r$ compared to $\gamma_2$ determines the equilibrium constant. The \textit{total} reaction rate $k = k_{f} + k_{r}$ is independent of $r$ because $k_{f} \sim r$ while $k_{r} \sim \gamma_2 \gg r$ so that $k \approx k_r$. That is, the system equilibrates on the time scale determined by $\gamma_2$, which is much faster than the time scale dictated by the pumping rate. The forward reaction rate is given by $kK/(1+K) \approx k K$, where the approximation follows from the fact that $K \approx r/\gamma_2 \ll 1$. Unlike the total reaction rate $k_f + k_r$, the forward reaction rate is directly proportional to $r$,
\begin{align}
k_f = \frac{r}{1+\gamma_2/\gamma_1} .
\end{align}
This is to be expected, since the generation of products is limited by the rate $r$ at which population can be transferred out of state $\ket{1}$ and into $\ket{3}$. This rate is proportional, but not equal, to $r$. For example, if $\gamma_1 = \gamma_2$, then $k  = r/2$. This reflects the fact that the forward reaction rate is also proportional to the fraction of the excited population that ends up in a product state. This quantity is known as the reaction yield and is determined by the ratio $\gamma_2 / \gamma_1$. Indeed, using the analytical solution with $\gamma_2 = \gamma_1$ and $r \ll \ \gamma_1$ shows that a single exponential governed by the rates obtained from $I_0$ provides a reasonable estimate of the dynamics (not shown). Other progress moments may also be evaluated to make the results increasingly accurate. 

The result of the above analysis may be generalized, as follows. At early times relative to $t_{\mathrm{chem}}$, the excited population of a system excited at rate $r$ is equal to $rt$. The product population is then given by $P(t) = Y(t) \hspace*{0.05cm}  rt$, where $Y(t)$ is the time-dependent reaction yield, equal to the fraction of product population relative to the total excited population. According to Eq. (\ref{eq:rate_law}), the product population satisfies $\partial_t P(t) = k_f \hspace*{0.1cm} \mathrm{exp}(-[k_f+k_r]t)$ at times $t>t^*$. Since $\mathrm{exp}(-[k_f+k_r]t^*) \approx 1$ for $t^* \ll [k_f+k_r]^{-1}$, the forward reaction rate is  $k_f \approx \partial_t P(t) \vert_{t=t^*} =  Y(t^*) \hspace*{0.05cm} r + [\partial_t Y(t)] rt \vert_{t=t^*}$.
When the rate of excitation is rate-limiting, the dynamics in the excited state manifold occur on a much shorter time scale than $r^{-1}$, and so the reaction yield is expected to stabilize at a time $t^* \ll t_{\mathrm{chem}}$. Therefore, the forward reaction rate is given by $k_f = Y(t^*) \hspace*{0.05cm} r$, and so the benefit of the progress moment approach is in obtaining $Y(t^*)$ without direct simulation to $t^*$.  
\subsection{$\boldsymbol{V}$-system}
Consider next a standard \cite{timur_prl}  three-level $V$-type system [Fig. \ref{fig:systems_schematic}(b)], where the transition frequency between the ground state $\ket{1}$ and the excited state $\ket{2}$ is on the order of the optical frequency $ \omega_0$, while the excited state splitting is given by $\Delta \ll \omega_0$. Each excited state $\ket{i}$ is dipole-connected to the ground state through an incoherent radiation field, leading to excitation of each state at rate $r_i$ and spontaneous emission at rate $\gamma_i$. Writing the density operator in vector form as $\lket{\rho} = [\rho_{11},\rho_{22},\rho_{33},\rho^R_{23},\rho^I_{23} ]^{T}$, where $\rho_{ij} = \bra{i} \hat{\rho} \ket{j}$, $R$ and $I$ denote the real and imaginary parts, respectively, and $T$ denotes matrix transposition, the Liouvillian is obtained within the second-order Born-Markov approximation as \cite{amr_v_system,timur_prl}
	\begin{align}
	\hat{\mathcal{L}} = 	\setlength{\arraycolsep}{3pt}  \begin{pmatrix} -2r & \gamma +r & \gamma +r  & 2 (\gamma +r) p & 0\\  r & -(\gamma +r)& 0  & -(\gamma+r) p & 0  \\  r & 0 & -(\gamma+r) &  -(\gamma +r)p & 0 \\ pr & - (\gamma+r) p /2 & - (\gamma +r)p /2 &-( \gamma+r) & \Delta \\ 0 & 0 & 0 &- \Delta &- (\gamma+r)   \end{pmatrix}.
	\end{align}
	 Here the dipole alignment factor is defined as $p= \boldsymbol{\mu}_{12} \cdot \boldsymbol{\mu}_{13} / \vert {\mu_{12}} {\mu_{13}}\vert$, where $\boldsymbol{\mu}_{ij} = \bra{i} \hat{\boldsymbol{\mu}} \ket{j} $ denotes the vector transition dipole matrix element between levels $\ket{i}$ and $\ket{j}$. The magnitude of the coherence generated by excitation is determined by the value of $p$. In simplifying the expression for $\hat{\mathcal{L}}$, we have set $\gamma_1 = \gamma_2 = \gamma$ and $r_1 = r_2 = r$. The secular approximation has been applied to the coherences between states $\ket{1}$ and $\ket{2}$ and states $\ket{1}$ and $\ket{3}$, decoupling the populations and the excited state coherences. However, the secular approximation has not been applied to the excited state coherences. The unique steady state is canonical, $\lket{\rho_s} = [\bar{n}+1, \  \bar{n}, \  \bar{n}, \ 0, \ 0]^{T}/(3\bar{n}+1)$, where we have defined the effective thermal occupation number as $\bar{n} = r/\gamma$ and written $\lket{\rho_s}$ in the vector form introduced above. For a system that is initially in the ground state, $\lket{\rho_0} = [1,0,0,0,0]^T$, the vector $\lket{\delta \rho_0}$ is obtained as
	\begin{align}
		\lket{\delta \rho_0} =  \frac{\gamma^{-1}(1+\Delta^2/\gamma^2)}{ (3\bar{n}+1)(-1+p^2-\Delta^2/\gamma^2)} 	\setlength{\arraycolsep}{3pt}  \bigg[ -2\bar{n}  , \  \bar{n}, \  \bar{n}, \  \frac{\bar{n}p}{1+\Delta^2/\gamma^2}  , \  \frac{\bar{n}p(\Delta/\gamma)}{1+\Delta^2/\gamma^2} \bigg]^{T}, \label{eq:A_Vsystem}
	\end{align}
	where we have used the fact that $\bar{n} \ll 1$ to simplify the expression.

	As an example, we consider the population of state $\ket{2}$ as a function of time. Hence, we wish to monitor the expectation value of the projector $\hat{O} = \ket{2}\bra{2}$ as a function of time. 
	The normalized zeroth progress moment is found to be
	\begin{align}
	 [k^{(0)}]^{-1} =  \lbraket{O \vert \delta \rho_0}/ \braket{\hat{\chi}(0)}   = \gamma^{-1} \frac{1+(\Delta/\gamma)^2}{ (1-p^2)+ (\Delta/\gamma)^2} \label{eq:prog0},
	\end{align}
	where $\lbraket{O \vert \delta \rho_0}$ denotes the inner product of vectors $\lket{O}$ and $\lket{\delta \rho_0}$ (equivalent to $\mathrm{Tr}[\hat{O} (\delta \hat{\rho}_0)]$ in Hilbert space), and we have once again made use of the fact that $\bar{n} \ll 1$ to simplify the expression. It is clear from Eq. (\ref{eq:prog0}) that the time scale of the population evolution can be expressed in units of $\gamma^{-1}$ in a way that is determined by the dimensionless parameters $p^2$ and $(\Delta/\gamma)^2$. Consider first the underdamped regime, $\Delta/\gamma \gg 1$ \cite{timur_prl}, in which the energy splitting is much larger than the spontaneous emission rate. Then for any choice of $p$, we have $1-p^2 +(\Delta/\gamma)^2 \approx (\Delta/\gamma)^2$, since $0 \leq p^2 \leq 1$. Then Eq. (\ref{eq:prog0}) simplifies as $k^{(0)} = \gamma$. Using only the zeroth progress moment to reconstruct the dynamics yields $\rho_{22}(t) = \braket{\hat{\chi}(t)} + \rho_{22}(\infty) = \braket{\hat{\chi}(0)} e^{-k^{(0)}t } + \bar{n}$, giving
	\begin{align}
	\rho_{22}(t) = \frac{r}{\gamma} \left(1- e^{-\gamma t} \right), \label{eq:underdamped}  \ \ \ \  \ \ (\mathrm{underdamped})
	\end{align}
	which is the exact analytical result in the underdamped limit \cite{timur_prl}. 
	
	Consider next the overdamped regime, i.e. $\Delta/\gamma \ll 1$. In the case $p=0$, the zeroth progress moment still gives $k^{(0)} = \gamma$, which yields Eq. (\ref{eq:underdamped}) once more. However, in the case of parallel or anti-parallel dipoles, $p^2 =1$, Eq. (\ref{eq:prog0}) simplifies as
	\begin{align}
	[k^{(0)}]^{-1} = (\gamma^2/\Delta^2 +1 )\gamma^{-1}\approx \left(\frac{\gamma^2}{ \Delta^2}\right) \gamma^{-1}  \gg \gamma^{-1}  \label{eq:overdamped}  \ \ \ \  \ \ (\mathrm{overdamped}) .
	\end{align}
	The associated value of $\rho_{22}(t)$ is then here estimated as $\rho_{22}(t) \approx (r/\gamma)[1-e^{-\Delta^2 t /\gamma}]$. By comparison, the exact result in this overdamped regime is \cite{timur_prl}
	\begin{align}
	\rho_{22}(t) = \frac{r}{\gamma} \left(  1- \frac{1}{2} e^{-2\gamma t} - \frac{1}{2}e^{-\Delta^2 t / 2 \gamma} \right)  \ \ \ \  \ \ (\mathrm{overdamped}) . \label{eq:overdamped1}
	\end{align}
	This result is comprised of two exponentials, each of which is weighted by a factor of $1/2$. One decays at the rapid rate $2 \gamma$, and the other at the slow rate $\Delta^2 /2\gamma \ll \gamma$. In this regime, quasi-stationary coherences were also found to survive on the long time scale $2\gamma/\Delta^2$ \cite{timur_prl}. Using only the zeroth progress moment and assuming exponential decay has given us a weighting factor of $1$ and the slow decay rate of $\Delta^2 /\gamma$. As will be discussed in section \ref{sec:numerics}, one does not know in advance how many progress moments are needed to accurately reconstruct the dynamics. It is therefore important to check that the reconstructed dynamics obtained from $n$ progress moments also agree with the progress moment $I_{n+1}$. In this example, evaluating the first progress moment indicates that the decay cannot be characterized by only a single exponential. That is,
	\begin{align}
	I_1 / \braket{\hat{\chi}(0)} \approx 2 \left(  \frac{\gamma^4}{ \Delta^4} \right) \gamma^{-2}\ = 2 \left(k^{(0)} \right)^{-2} \label{eq:prog1}  \ \ \ \  \ \ (\mathrm{overdamped}) .
	\end{align}
	Since an ideal exponential decay would satisfy $I_1/\braket{\hat{\chi}(0)} = \left(k^{(0)} \right)^{-2} $, the factor of two that multiplies the second progress moment indicates that the dynamics at long times are far slower than the dynamics at short times. This in turn indicates that there are at least two disparate time scales that characterize the decay of the progress variable. 
	
	In obtaining the approximate progress moments in Eqs. (\ref{eq:overdamped}) and (\ref{eq:prog1}), we have ignored small terms that are of order $(\Delta/\gamma)^2$ and  $(\Delta/\gamma)^4$, respectively. If these terms are included, then \textit{analytically} solving Eq. (\ref{eq:basis_solve}) for the $f_m$ and $k_m$ yields $f_0 = f_1 =\braket{ \hat{\chi}(0)}/2$, $k_0 = \Delta^2/2\gamma$ and $k_1  = 2\gamma$. This is in exact agreement with the result of Eq. (\ref{eq:overdamped1}), since the analytical result contains only two exponentials. However, if $I_0$ and $I_1$ were calculated \textit{numerically}, the small corrections to Eqs. (\ref{eq:overdamped}) and (\ref{eq:prog1}) could be too small to be accurate. Numerical tests indicate that the calculated values of $f_0$, $f_1$ and $k_0$ are robust to small changes in $I_0$ and $I_1$, but that the calculated value of $k_1$ is not. In each case $k_1$ is always found to satisfy $k_1 \gg k_0$, but its exact value is sensitive to small changes in $I_0$ and $I_1$.
If detailed information about early time dynamics is of interest, then it is useful to apply a Laplace transform to the progress variable, $\int_{0}^{\infty} dt \braket{\hat{\chi}(t)} e^{-k_F t}$, at a set of fast rates $k_F$, as described in
Appendix \ref{app:laplace}. This approach would yield a more accurate estimate of $k_1$.
\section{Numerical implementation}
\label{sec:numerics}
This section provides details of the computational issues addressed and resolved in implementing this approach. A molecular example is discussed in section \ref{sec:pyrazine}.
\subsection{Solution of $\boldsymbol{\hat{\rho}_s}$ and $\boldsymbol{\delta \hat{\rho}_n}$}
\label{subsec:solution}
The main computational challenge of the approach introduced here is to solve for $\hat{\rho}_s$ and the $\delta \hat{\rho}_n$. Below we show how $\hat{\rho}_s$ and the $\delta \hat{\rho}_n$ can be calculated through an iterative method. The method involves computing $\hat{\mathcal{L}}\hat{\rho}_s^{(i)}$ and $\hat{\mathcal{L}}[\delta \hat{\rho}_n^{(i)}]$, where the $i$ superscript denotes the $i$th iterative approximation to the operator of interest. By contrast, the solution of the master equation would require calculating $\hat{\mathcal{L}} \hat{\rho}(t)$. Both involve calculating the effect of $\hat{\mathcal{L}}$. In principle, solving for $\hat{\rho}_s$ and the $\delta \hat{\rho}_n$ could take as long as propagating $\hat{\mathcal{L}} \hat{\rho}(t)$ over the time scale of interest. However, in this work we found that solving for the operators with a secular preconditioner is far faster than computing the dynamics over the relevant time scales (see section \ref{sec:pyrazine} and section \ref{subsec:solution} below).

Referring to $\delta \hat{\rho}_n$ and $\hat{\rho}_s$ with the generic label $\hat{X}$, each problem can be written as
\begin{align}
 \hat{X} = [\hat{\mathcal{L}} + \hat{\kappa} ]^{-1} \hat{v}, \label{eq:x_eq}
\end{align}
where $\hat{v}$ is given and $\hat{\kappa}$ enforces a mathematical constraint that renders $(\hat{\mathcal{L}} + \hat{\kappa})$ non-singular. For example, the steady state density operator can be written as \cite{steady_methods}
\begin{align}
\hat{\rho}_s = [\hat{\mathcal{L}} +w \hat{\mathcal{T}} ]^{-1} \hat{w}, \label{eq:rho_s}
\end{align}
where $\mathcal{T}_{ijkl} = \delta_{kl} \delta_{ij} \delta_{i1}$ and $w_{ij} =w \delta_{ij} \delta_{i1}$ for any constant $w$. As noted above, the Liouvillian contribution enforces the condition $\hat{\mathcal{L}}\hat{\rho}_s = \hat{0}$. The superoperator $\hat{\mathcal{T}}$ ensures $\mathrm{Tr}[\hat{\rho}_s] =1$, through $w\hat{\mathcal{T}} \hat{\rho}_s = w \sum_{kl} \mathcal{T}_{ijkl} (\rho_{s})_{kl} \ket{i} \bra{j} = \mathrm{Tr}[\hat{\rho}_s]\hat{w}$. Similarly, for $\delta \hat{\rho}_n$ we have
\begin{align}
\delta \hat{\rho}_n =- [ \hat{\mathcal{L}} + w \hat{\mathcal{T}} ]^{-1} \hat{c}_n,
\end{align}
whereas in the case of a Laplace transform, $\hat{\kappa}$ is proportional to the identity operator \cite{jesenko}. 

Equation (\ref{eq:x_eq}) is a linear equation in the unknown $\hat{X}$ of the form $\hat{F} \hat{X} = \hat{v}$, where $\hat{F} = \hat{\mathcal{L}} + \hat{\kappa}$. This equation can be solved iteratively by writing \cite{matrix_book} $\hat{F}= \hat{G}+\hat{J}$, where $\hat{G}$ is known as a preconditioner and approximates $\hat{F}$, and where the contribution of $\hat{J}$ is ``small.'' If $\hat{G}^{-1}$ is known, then the linear equation can be solved by iterating \cite{matrix_book}
\begin{align}
\hat{X}^{(l+1)} = \hat{G}^{-1}(\hat{v}- \hat{J}\hat{X}^{(l)}). \label{eq:iteration}
\end{align}
Here, the superscript $l$ indicates the $l$th iterative approximation to $\hat{X}$. This algorithm converges to the true solution provided that $\hat{H}$ is small compared to $\hat{G}^{-1}$, i.e. such that \cite{matrix_book}
\begin{align}
S(\hat{G}^{-1}\hat{J}) \equiv \mathrm{max} \{ \vert \lambda(\hat{G}^{-1}\hat{J}) \vert \} < 1. \label{eq:convergence}
\end{align}
Here, $S(\hat{G}^{-1}\hat{J})$ denotes the spectral radius of $\hat{G}^{-1}\hat{J}$, which is the largest absolute value of the eigenvalues $\lambda$ of $\hat{G}^{-1}\hat{J}$. In general, the success of an iterative algorithm depends crucially on the choice of preconditioner. A review and comparison of common preconditioners that can be used to solve Eq. (\ref{eq:rho_s}) can be found in Refs. \cite{steady_methods} and \cite{nation_iterative}. In this work, we propose that here, for a physical system coupled to multiple thermal baths, the rate law description associated with the secular approximation should provide a reasonable description of the dynamics. Therefore, a natural decomposition of $\hat{F}$ is into its secular and non-secular components:
\begin{align}
\hat{G} = \hat{\mathcal{L}}_{\mathrm{sec}} + \hat{\kappa}_{\mathrm{sec}}, \ \ \ \, \hat{J} = \hat{\mathcal{L}}_{\mathrm{ns}} + \hat{\kappa}- \hat{\kappa}_{\mathrm{sec}}.
\end{align}
Here, $\hat{\mathcal{L}}_{\mathrm{sec}}$ is the Liouvillian associated with the secular approximation, and $\hat{\mathcal{L}}_{\mathrm{ns}} = \hat{\mathcal{L}} -\hat{\mathcal{L}}_{\mathrm{sec}}$. The subscripts on $\kappa$ are similarly defined. The superoperator $w\hat{\mathcal{T}}$ is in fact a secular operator, since its input is the diagonal elements of its argument, and its output is also diagonal. The identity operator is also secular, and so in both cases $\hat{\kappa} = \hat{\kappa}_{\mathrm{sec}}$. However, to keep the analysis general, we explicitly resolve $\hat{\kappa}$ into its secular and non-secular components.

In the energy eigenstate basis $\hat{H}_{\mathrm{S}}\ket{n} = E_n \ket{n}$, the effect of $\hat{\mathcal{L}}_{\mathrm{sec}}$ is given by $\bra{n}\hat{\mathcal{L}}_{\mathrm{sec}}\hat{p}\ket{m} = -i[ \omega_{nm} + \gamma_{nm}] p_{nm}$ for $n\neq m$, where $\hat{p}$ is an arbitrary operator, $\omega_{nm} = (E_n-E_m)/\hbar$ is the transition frequency, and $\gamma_{nm}$ is the decoherence rate. For this reason, inverting $\hat{G}$ is straightforward. In particular, writing $\hat{G}^{-1} \hat{v} \equiv \hat{p}$, we have
\begin{align}
& [\tilde{\mathcal{L}} + \tilde{\kappa} ] \tilde{p} = \tilde{v}, \label{eq:diag_inverse} \\
& p_{nm} = -\frac{v_{nm}}{i \omega_{nm} + \gamma_{nm}}, \ \ \ n \neq m. \label{eq:odiag_inverse}
\end{align}
Here, a tilde denotes a restriction to the subspace spanned by the diagonal elements of $\hat{p}$ and $\hat{v}$---that is, $\tilde{\mathcal{L}}_{nm} \equiv \mathcal{L}_{\mathrm{sec},nnmm}$, $\tilde{p}_{m} \equiv p_{mm}$, and similarly for $\tilde{\kappa}$ and $\tilde{v}$. 
 Equation (\ref{eq:diag_inverse}) thus defines a matrix equation for the $d$ diagonal elements of $p$. For model pyrazine, discussed as an example in section \ref{sec:pyrazine}, $d= 660$. Linear equations for 660 variables can be solved in less than a second, where discussion of computational efficiency here and below is with respect to a laptop with a 2.9 GHz processor. Equation (\ref{eq:odiag_inverse}) involves $d^2 - d$ components and is explicitly solved for in terms of the frequencies and decoherence rates. The prefactor multiplying $v_{nm}$ can be stored as a matrix of size $d \times d$, and $p$ can thus be obtained through element-wise multiplication of this matrix with $v$. This operation is performed in a fraction of a second using typical numerical software such as Matlab \cite{matlab}. The inverse $\hat{G}^{-1} \hat{v}$ can therefore be calculated rapidly for any $\hat{v}$.

If the non-secular contribution to $\delta \hat{\rho}_n$ or $\hat{\rho}_s$ is sufficiently small, then the inequality (\ref{eq:convergence}) will be satisfied, and both operators can be obtained by iterating Eq. (\ref{eq:iteration}). Since $\hat{G}^{-1}$ can be calculated as described above, this process efficiently solves the linear equations of interest. However, if the non-secular contribution is significant, the iterative algorithm may not converge. In this case, we instead ``scale down'' the non-secular contribution by a factor $0 < \eta <1$, through
\begin{align}
\hat{G} = \hat{\mathcal{L}}_{\mathrm{sec}} + \hat{\kappa}_{1-\eta} + (1-\eta) \hat{\mathcal{L}}_{\mathrm{ns}}, \ \ \ \hat{J} = \eta \hat{\mathcal{L}}_{\mathrm{ns}} + \hat{\kappa} - \hat{\kappa}_{1-\eta}. \label{eq:1_minus_eta}
\end{align}
Here, $\hat{\kappa}_{1-\eta}$ is defined with respect to $\hat{\mathcal{L}}_{1-\eta}$; this notation is explained in the footnote \cite{ns_footnote}. Convergence of an iterative algorithm involving $\hat{G}$ and $\hat{J}$ defined by Eq. (\ref{eq:1_minus_eta}) is guaranteed if
\begin{align}
\eta S( \hat{G}^{-1}  \hat{\mathcal{L}}_{\mathrm{ns}} ) <1. \label{eq:scale_down_1}
\end{align}
The operator $\hat{G}$ has now gained a contribution $(1-\eta)\hat{\mathcal{L}}_{\mathrm{ns}}$, while $\hat{J}$ has lost this contribution. Therefore, for small enough $\eta$, the convergence condition is guaranteed to be met. The inverse of $\hat{G}$ itself can then obtained by resolving it into two further components:
\begin{align}
& \hat{G} = \hat{G}_1 + \hat{G}_2,  \ \ \ \ \hat{G}_1 = \hat{\mathcal{L}}_{\mathrm{sec}} + \hat{\kappa}_{\mathrm{sec}}, \nonumber \\ &\hat{G}_2 = (1-\eta)\hat{\mathcal{L}}_{ns} + \hat{\kappa}_{1-\eta} - \hat{\kappa}_{\mathrm{sec}}.
\end{align}
$\hat{G}^{-1}$ is then obtained by calculating $\hat{G}_1^{-1}$ and iterating the contribution from $\hat{G}_2$. Convergence is guaranteed if
\begin{align}
(1-\eta) S(\hat{G}_{1}^{-1} \hat{\mathcal{L}}_{\mathrm{ns}}) < 1. \label{eq:convergence2}
\end{align}
If there exists an $\eta$ such that the inequalities (\ref{eq:scale_down_1}) and (\ref{eq:convergence2}) are satisfied, then $\hat{F}^{-1} \hat{v}$ can be calculated iteratively. This is indeed the case if the contribution of $\hat{\mathcal{L}}_{\mathrm{ns}}$ is fairly small compared to that of $\hat{\mathcal{L}}_{\mathrm{sec}}$. The iteration scheme is thus summarized as follows:
\begin{enumerate}
	\item[] The operator $\hat{X}\equiv[ \hat{\mathcal{L}}+ \hat{\kappa}]^{-1} \hat{v}$ is obtained by iterating over the $l$th approximation to $\hat{X}$, denoted with a superscript by $\hat{X}^{(l)}$. To evaluate $ \hat{X}^{(l+1)}$ given $\hat{X}^{(l)}$, choose an initial guess $\hat{X}^{(l+1)}_{0}$ (a reasonable choice is $\hat{X}_{\mathrm{sec}}$) and iterate over $m$ for a given $\eta$:
	\begin{align}
	&\hat{X}^{(l+1)} = \lim_{m\to\infty} \hat{X}_{m+1}^{(l+1)} = [\hat{\mathcal{L}}_{\mathrm{sec}} + \hat{ \kappa}_{\mathrm{sec}}   ]^{-1}\nonumber \\
	& [ \hat{v} - \eta \hat{\mathcal{L}}_{\mathrm{ns}} \hat{X}^{(l)} -(\hat{\kappa}-\hat{\kappa}_{1-\eta})\hat{X}^{(l)} \nonumber \\
	&\indent \indent - (1-\eta) \hat{\mathcal{L}}_{\mathrm{ns}} \hat{X}_{m}^{(l+1)}   - (\hat{\kappa}_{1-\eta}-\hat{\kappa}_{\mathrm{sec}})\hat{X}_m^{(l+1)}].
	\end{align}
	\item[] The effect of $[\hat{\mathcal{L}}_{\mathrm{sec}} + \hat{\kappa}_{\mathrm{sec}}]^{-1}$ can be calculated using Eqs. (\ref{eq:diag_inverse}) and (\ref{eq:odiag_inverse}).
\end{enumerate}
In principle this algorithm can be further extended if the non-secular contribution is large enough that no $\eta$ can be found such that convergence is obtained. In this case, $\mathcal{L}_{\mathrm{sec}} + \hat{\kappa}_{1-\eta} + (1-\eta) \hat{\mathcal{L}}_{\mathrm{ns}}$ can be further resolved into $\hat{\mathcal{L}}_{\mathrm{sec}} + \hat{\kappa}_{1-\eta-\varepsilon} + (1-\eta -\varepsilon) \hat{\mathcal{L}}_{\mathrm{ns}}$ and $\varepsilon \hat{\mathcal{L}}_{\mathrm{ns}} + \hat{\kappa}_{1-\eta}  - \hat{\kappa}_{1-\eta-\varepsilon}$, for some $\varepsilon <1$. The inverse $[\hat{\mathcal{L}}_{\mathrm{sec}} + \hat{\kappa}_{\mathrm{sec}} + (1-\eta-\varepsilon) \hat{\mathcal{L}}_{\mathrm{ns}}]^{-1}$ would then be calculated iteratively. In practice we have not found this to be necessary. 

In section \ref{sec:pyrazine} we discuss model pyrazine, as well as modifications to the model pyrazine system that result in a moderate non-secular contribution. Yet even in this case, the algorithm described above converges for a wide range of $\eta$ between $0.2$ and $0.7$. 
Interestingly, the semi-analytical approach to calculating $\hat{\rho}_s$ presented in Refs. \cite{red_steady1,red_steady2,red_steady3} fails in this case because of the near-degeneracy of several eigenstates. That is, the assumption that the energy splitting is much greater than the system-bath coupling is not fulfilled, and the semi-analytical approach cannot be used. When the assumption is valid, however, the algorithm that we use can be applied without resolving $\hat{\mathcal{L}}_{\mathrm{ns}}$ into two components. In this case it can be regarded as a simple extension of the approach used in Ref. \cite{red_steady3} to higher order in perturbation theory.

This algorithm is straightforward to implement, provided that the effect of $\hat{\mathcal{L}}$ can be easily calculated. This is indeed the case for the full non-secular Redfield equations of motion, wherein $\hat{\mathcal{L}}\hat{v}$ can be calculated via matrix multiplication of matrices of size $d\times d$ \cite{pollard}. It is also the case for the Lindblad equations of motion. In the case of model pyrazine described below, even when using modified system parameters to enhance the non-secular contribution, excellent convergence for such an equation is obtained 1-3 minutes. 

We conclude this subsection by mentioning three useful considerations when implementing this algorithm. First, when taking the diagonal component of $\hat{v}$ to solve $[\hat{\mathcal{L}}_{\mathrm{sec}} + \hat{\kappa}_{\mathrm{sec}}]^{-1}\hat{v}$, it is advantageous to consider only the real part. Since every operator considered should be Hermitian, the diagonal component should already be real. In practice, however, there may be a small imaginary component introduced by roundoff error. Removing this imaginary component stabilizes the algorithm, ensures the Hermiticity of the operators and enhances convergence. Second, some care is required when checking convergence, especially for $\hat{\rho}_s$. Depending on the particular problem, the precision of $\hat{\rho}_s$ may be more or less important. For example, in the internal conversion of pyrazine described below, it is vital that the elements of $\hat{\rho}_s$ for energies above the minimum of the $S_1$ potential well are accurate. When these elements, crucial for describing the process of interest, are small in magnitude compared to other elements it is important that they are converged to a higher accuracy. Third, it is important to remember that $w$ is a dimensioned constant with units of inverse time. Therefore, to ensure a stable iterative scheme, one should choose this parameter such that it is of the same order of magnitude as typical system transition rates. For model pyrazine, for example, setting $w^{-1}$ equal to $2.4$ fs yielded a stable algorithm. In practice this is straightforward to implement through trial and error, since divergences in the algorithm become apparent after only a few iterations.
\subsection{Projection onto exponential basis}
\label{subsection:projection}
Once $\hat{\rho}_s$ and $\delta \hat{\rho}_n$ are obtained, the progress variable can be reconstructed by solving Eq. (\ref{eq:basis_solve}). Doing so is straightforward. However, there are several subtle theoretical and computational aspects deserving of discussion.

First, we must consider whether to treat Eq. (\ref{eq:basis_solve}) as a system of linear equations, with the rates $k_m$ chosen \textit{a priori}, or as a nonlinear system that should be solved for both the amplitudes $f_m$ and rates $k_m$. If the first approach is taken then, in principle, the number of basis functions $m_{\mathrm{max}}+1$ is unbounded. Therefore, one is always free to use more basis functions than values of $I_n$ and $\braket{\hat{\chi}(0)}$. As a consequence, it is always possible to create an underdetermined system and thus choose at least one parameter arbitrarily while exactly satisfying Eq. (\ref{eq:basis_solve}). Similarly, choosing $m_{\mathrm{max}} = n_{\mathrm{max}}$ ensures an exact solution for $n \leq n_{\mathrm{max}}$. 
However, such solutions will not satisfy Eq. (\ref{eq:basis_solve}) for $n > n_{\mathrm{max}}$. In fact, the larger the number of arbitrary parameters introduced, the more severe the disagreement for $n>  n_{\mathrm{max}}$. Therefore, to obtain an accurate projection, it is necessary to choose $m_{\mathrm{max}} < n_{\mathrm{max}}$, so that the system is overdetermined. The $f_m$ are then chosen as the best fit parameters. The goodness of fit and the agreement for $n > n_{ \mathrm{max} }$ are used to assess the accuracy of the reconstructed function.

If the second approach is taken, then one can numerically solve Eq. (\ref{eq:basis_solve}) for the $f_m$ and $k_m$. Since the number of basis functions is $m_{\mathrm{max}}+1$, the system is exactly determined for $n_{\mathrm{max}}= 2m_{\mathrm{max}}+1$. In general the solutions to these equations can be complex, yet we require all variables to be real and the $k_m$ to be positive.
In practice, however, the solutions are often real and the $k_m$ positive when $m_{\mathrm{max}}$ is small enough. That is, there often exists some $\mathcal{M}$, such that for $m_{\mathrm{max}} + 1\leq \mathcal{M}$, the solutions are real and the $k_m$ positive. For $m_{\mathrm{max}} +1> \mathcal{M}$, the solutions become complex, and/or the $k_m$ negative. The simplest case is $\mathcal{M}=1$, which has the solution $f_0 = \braket{\hat{\chi}(0)}$ and $k_0^{-1} = I_0/ \braket{\hat{\chi}(0)}$, where $k_0$ is positive if the first progress moment has the same sign as $f_0$.
For model pyrazine below, we have found that in many cases $\mathcal{M}=3$. The solution to the nonlinear equations may be judged according to its predicted progress moments for $n> n_{\mathrm{max}}$. If the solution agrees with higher order progress moments that were not included in its construction, then it is deemed an accurate solution. If not, one can instead obtain a best nonlinear fit for the $f_m$ and $k_m$ using $n_{\mathrm{max}} > 2m_{\mathrm{max}} + 1$.

In general, treating both the $f_m$ and $k_m$ as unknowns is more fruitful, since choosing the basis functions arbitrarily may require a larger number of basis functions for convergence. One approach to obtaining the $f_m$ and $k_m$ is to solve the nonlinear equations with an increasing number of basis functions until the solutions become complex. The number of basis functions for which the solutions are complex define the value of $\mathcal{M}$. The $f_m$ and $k_m$ are then taken to be the solutions obtained for $m_{\mathrm{max}}+1= \mathcal{M}$, which we have found to be accurate in many cases. When a nonlinear fit is required, the ``NonlinearModelFit'' option in Mathematica \cite{Mathematica} can be used, as follows. 
Since the $k_m$ are expected to vary over many orders of magnitude, we re-write Eq. (\ref{eq:basis_solve}) in terms of $\tilde{k}_m \equiv \mathrm{ln}(k_m)$:
\begin{align}
&\sum_{m=0}^{m_{\mathrm{max}}} f_m e^{-n \cdot \tilde{k}_m }= y_n, \label{eq:basis_solve2}
\end{align}
where $y_n$ is defined as the right-hand side of Eq. (\ref{eq:basis_solve}). We have found that the narrower distribution of $\tilde{k}_m$ values improves convergence. 
Next, note that typical solvers minimize the sum $\sum_n (\mathrm{lhs}_n -\mathrm{rhs}_n)^2$ or a similar parameter, where $\mathrm{lhs}_n$ is the left-hand side and $\mathrm{rhs}_n$ is the right-hand side of Eq. (\ref{eq:basis_solve}) with index $n$. In Eq. (\ref{eq:basis_solve2}), $\mathrm{lhs}_n = \sum_{m} f_m e^{-n \cdot \tilde{k}_m }$ and $\mathrm{rhs}_n =  y_n$. Depending on the chosen units, the $y_n$ will either tend to zero or infinity for large $n$ \cite{decay_vs_growth}. Therefore, larger values of $n$ will be given either too much weight or too little. To correct this, we assign weights to each value of $y_n$, through $W_n = \vert y_n \vert^{-b}$, where $W_n$ is the $n$th weight, and $b$ is a parameter that can be varied. This can be implemented using the ``Weights'' function in Mathematica \cite{Mathematica}. 
We have found that $b=2$ is typically a good choice. 

Equation (\ref{eq:basis_solve2}) can be solved numerically given an initial seed value for the $f_m$ and $k_m$. A reasonable set of seed values is the set obtained from the exact numerical solution for $m_{\mathrm{max}} +1 = \mathcal{M}$ \cite{footnote}.
Both procedures associated with the second approach were successfully used to recreate the progress variable for model pyrazine below using only five progress moments and three basis functions.
\section{Internal conversion of model pyrazine governed by Redfield dynamics}
\label{sec:pyrazine}
As a numerical example, consider the internal conversion of model pyrazine driven by incoherent light. The process of interest is incoherent excitation from the $S_0$ ground electronic manifold to the $S_1/S_2$ excited manifold, and the associated bath-induced decay to $S_1/S_2$ states of lower energy. Following Ref. \cite{pyrazine_2s2m} we adopt a minimal model of pyrazine consisting of the three diabatic electronic states ($S_0$, $S_1$ and $S_2$) with two vibrational modes. The remaining modes of the molecule, assumed to be harmonic and only affected by excitation indirectly, are considered as a bath and treated within Redfield theory.
\subsection{Master equation}
 The electronic states consist of the ground $S_0$ state as well as the two excited $S_1$ and $S_2$ states, while the vibrational modes comprise a harmonic tuning mode (frequency $\omega_t$) and a harmonic coupling mode (frequency $\omega_c$). In the limit of linear electronic-vibrational coupling, the system Hamiltonian is given by \cite{pyrazine_2s2m}
\begin{align}
& \hat{H}_{\mathrm{S}} = \sum_{k=0}^{2} \ket{\phi_k}  \bra{\phi_k} \hat{h}_k + \lambda \hat{x}_c (\ket{\phi_1}\bra{\phi_2} + \ \mathrm{h.c.}) \label{eq:hamiltonian} \\
& \hat{h}_k = \hat{h}_0   + \kappa_k \hat{x}_t + E_k \\
& \hat{h}_0 = \hbar \omega_c \bigg(\hat{a}_c^{\dagger}  \hat{a}_c+ \frac{1}{2} \bigg) + \hbar \omega_t \bigg(\hat{a}_t^{\dagger}  \hat{a}_t+ \frac{1}{2} \bigg).
\end{align}
Here, $\ket{\phi_0}$ is the ground electronic state, $\ket{\phi_1}$ and $\ket{\phi_2}$ are the diabatic excited electronic states, $\hat{x}_c = (\hat{a}_c+\hat{a}^{\dagger}_c)/\sqrt{2}$ is the position operator for the system coupling mode, and $\hat{x}_t = (\hat{a}_t+\hat{a}^{\dagger}_t)/\sqrt{2}$ is the position operator for the system tuning mode. $\hat{a}^{\dagger}$ and $\hat{a}$ are system mode creation and annihilation operators, respectively, $\lambda$ quantifies the vibronic coupling between the excited states, and h.c. denotes the Hermitian conjugate. The ground state component of the Hamiltonian is $\hat{h}_0$, which consists of two uncoupled harmonic oscillators centered at $\hat{x}_c, \ \hat{x}_t = \hat{0}$.  The excited states are characterized by excitation energies $E_k$ and intra-state electronic-vibrational coupling constants $\kappa_k$. 

The effect of the unreactive modes and the condensed phase environment are incorporated through bilinear coupling to a set of infinite harmonic oscillators, through \begin{align}
&\hat{H}_{\mathrm{B}} = \{ \ket{\phi_1}\bra{\phi_1}  +  \ket{\phi_2}\bra{\phi_2} \} \sum_n \hbar \omega_n \bigg(\hat{b}^{\dagger}_n \hat{b}_n + \frac{1}{2} \bigg) \\
& \hat{H}_{\mathrm{SB}} = \{ \ket{\phi_1}\bra{\phi_1}  +  \ket{\phi_2}\bra{\phi_2} \} \bigg( \sum_{nm}  g^{(n)}_{m} (\hat{a}_m + \hat{a}_m^{\dagger}) (\hat{b}_n + \hat{b}_n^{\dagger}) \bigg),
\end{align}
where B and SB refer to bath and system-bath, respectively, $n$ enumerates the bath degrees of freedom, $m=c, \ t$ enumerates the system coupling operators, and $g_m^{(n)}$ quantifies the system-bath coupling strength. The creation and annihilation operators of the $n$th bath mode are $\hat{b}_n^{\dagger}$ and $\hat{b}_n$. The system-bath coupling is characterized by the spectral density $J_m(\omega) = 2\pi \sum_n \vert g_m^{(n)} \vert^2 \delta(\omega - \omega_n)$ and the system coupling operators $\hat{a}_m+\hat{a}_m^{\dagger}$. The interaction of the system with the incoherent radiation field is given by \cite{breuer}
\begin{align}
&\hat{H}_{\mathrm{rad}} = \sum_{\mathbf{k}, \lambda} \hbar \omega_{k} \bigg(\hat{c}^{\dagger}_{\mathbf{k}, \lambda} \hat{c}_{\mathbf{k}, \lambda} + \frac{1}{2} \bigg) \\
& \hat{H}_{\mathrm{S-rad}} =  -i \boldsymbol{\mu} \cdot \sum_{\mathbf{k}, \lambda} \bigg( \frac{\hbar \omega_k }{2 \varepsilon_0 V} \bigg)^{1/2} \boldsymbol{\epsilon}_{\mathbf{k}, \lambda} \bigg(\hat{c}_{\mathbf{k}, \lambda}- \hat{c}^{\dagger}_{\mathbf{k}, \lambda} \bigg).
\end{align}
Here, rad denotes the radiation field, $\omega_k$ is the frequency of the $k$th electric field mode, $\mathbf{k}$ is the associated wavevector and $\lambda$ is the polarization, $c^{\dagger}_{\mathbf{k}, \lambda}$ is the mode creation operator, and $c_{\mathbf{k}, \lambda}$ is the mode annihilation operator. In the dipole approximation, the system-radiation field interaction Hamiltonian $H_{\mathrm{S-rad}}$  is characterized by the system dipole operator $\hat{\boldsymbol{\mu}}$, with $V$ being the cavity volume, $\varepsilon_0$ being the permittivity of free space, and $\boldsymbol{\epsilon}_{\mathbf{k}, \lambda}$ being the mode polarization vector. The total Hamiltonian is given by $\hat{H}_{\mathrm{tot}} = \hat{H}_{\mathrm{S}} + \hat{H}_{\mathrm{B}} + \hat{H}_{\mathrm{rad}} + \hat{H}_{\mathrm{SB}} + \hat{H}_{\mathrm{S-rad}}$. 

Taking the limit $V \to \infty$, applying a second-order Born-Markov approximation to both the bath and the radiation field, and tracing over the environment degrees of freedom yields the Redfield master equation \cite{hoki,timur2,breuer}
\begin{align}
\frac{d}{dt} \hat{\rho}= -\frac{i}{\hbar }[\hat{H}_{\mathrm{S}}, \hat{\rho}] + \sum_{kl} R_{ijkl} \rho_{kl} \ket{i} \bra{j}, \label{eq:master_nonsec}
\end{align}
where $\hat{H}_{\mathrm{S}}\ket{i}  = E_i \ket{i}$ enumerates the $i$th energy eigenstate. The Redfield tensor elements are given by \cite{psbr}
\begin{align}
&R_{ijkl} =\sum_{\alpha} \bigg( \Gamma^{(+)\alpha}_{ljik} + \Gamma^{(-)\alpha}_{ljik}  - \delta_{lj} \sum_r  \Gamma^{(+)\alpha}_{irrk} -\delta_{ik} \sum_r \Gamma^{(-)\alpha}_{lrrj} \bigg), \label{eq:redfield}
\end{align}
where $\alpha$ enumerates the system operators $\hat{q}^{\alpha}$ that are each coupled to an independent environment. The elements of the relaxation tensor associated with independent bath $\alpha$ are given by \cite{timur2}
\begin{subequations} \label{eq:gamma_tensor}
\begin{align}
& \Gamma^{(\pm)\alpha}_{ijkl}  = \frac{1}{2\pi} q^{\alpha}_{ij} q^{\alpha}_{kl} B^{(\pm)\alpha}_{kl} \\
&  B^{(+)\alpha}_{kl} = \begin{cases}
\pi J_{\alpha}(\omega_{kl}) \bar{n}_{\alpha}(\omega_{kl}), & \text{if } \omega_{kl}>0\\
\pi J_{\alpha}(-\omega_{kl}) (1+\bar{n}_{\alpha}(-\omega_{kl})), & \text{if } \omega_{kl}<0,
\end{cases}
\end{align}
\end{subequations}
where $B^{(-) \alpha}_{kl} = (B^{(+) \alpha}_{lk})^*$, $\bar{n}_{\alpha}(\omega) = [\mathrm{exp}(\hbar \omega / k_{\mathrm{B}} T_{\alpha} )-1]^{-1}$ is the thermal occupation number associated with the temperature of bath $\alpha$, $T_{\alpha} =300$ K, and $k_{\mathrm{B}}$ is Boltzmann's constant. The form of the radiative relaxation tensor takes a similar form \cite{psbr}:
\begin{align}
\Gamma^{(+)\mathrm{rad}}_{ijkl} = \frac{\vert \omega_{kl} \vert^3 }{6\pi \varepsilon_0 \hbar c^3} ( \boldsymbol{\mu}_{ij} \cdot \boldsymbol{\mu}_{kl}  ) \cdot  \begin{cases}
\bar{n}_{\mathrm{rad}}(\omega_{kl}), & \text{if } \omega_{kl}>0\\
1+\bar{n}_{\mathrm{rad}}(-\omega_{kl}),  &\text{if } \omega_{kl}<0,
\end{cases}
\label{eq:gamma_tensor_sun}
\end{align}
where boldface denotes a vector in real space and $\bar{n}_{\mathrm{rad}}(\omega)$ is defined with respect to the surface temperature of the sun, $T= 5800$ K. We assume that the system is directly illuminated by solar radiation unless otherwise noted; that is, the $C$ value describing filtering of the incident light \cite{hoki,timur2} is set to $1$. 

In the diabatic representation, the system dipole operator is given by $\hat{\boldsymbol{\mu}} = \hat{\boldsymbol{\mu}}_{01}(\hat{x}_c, \hat{x}_t) \ket{\phi_0}\bra{\phi_1} + \hat{\boldsymbol{\mu}}_{02}(\hat{x}_c, \hat{x}_t) \ket{\phi_0}\bra{\phi_2} + \mathrm{h.c.}$, where $\hat{\boldsymbol{\mu}}_{nm} = \bra{\phi_m}\hat{\boldsymbol{\mu}} \ket{\phi_n}$ Results of \textit{ab initio} calculations indicate that the $\hat{\boldsymbol{\mu}}_{nm}$ depend only weakly on the nuclear coordinates, and hence that the Franck-Condon approximation is valid \cite{pyrazine_ab_initio}. The dipole operator obtained from Ref. \cite{pyrazine_ab_initio} is given by $\hat{\boldsymbol{\mu}}_{01}= {{\mu}}_{01} \boldsymbol{x}$ and $\hat{\boldsymbol{\mu}}_{02}= {{\mu}}_{02} \boldsymbol{y}$, where $\boldsymbol{x}$ and $\boldsymbol{y}$ are orthogonal unit vectors in real space, and $\mu_{01} \approx 0.905$ D, $\mu_{02} \approx 1.575 $ D. It is clear that $\Gamma^{(+)\mathrm{rad}}$ may be resolved into two independent components characterized by the two coupling operators $\hat{\mu}^{(x)}$ and $\hat{\mu}^{(y)}$. Hence the coupling to the radiation field can be written in the form of Eqs. (\ref{eq:redfield}) and (\ref{eq:gamma_tensor}), with the radiation degree of freedom replaced by two independent degrees of freedom $\alpha$, with coupling operators $\hat{q}^{\alpha} = \{ \mu_{01} \ket{\phi_0} \bra{\phi_1} + \mathrm{h.c.}, \  \mu_{02} \ket{\phi_0} \bra{\phi_2} + \mathrm{h.c.} \}$. The associated spectral densities are identical, $J(\omega) = \omega ^3 /3\pi \varepsilon_0 \hbar c^3$, and the thermal occupation number is defined with respect to $T= 5800$ K. The separable form of Eq. (\ref{eq:gamma_tensor}) then allows Eq. (\ref{eq:master_nonsec}) to be written as a sum of matrix multiplications, through \cite{timur2,pollard}
\begin{align}
\hat{R} \hat{\rho} = -\hat{M}^{+}\hat{\rho} - \hat{\rho} \hat{M}^{-} + \sum_{\alpha} (\hat{P}^{(+)\alpha} \hat{\rho} \hat{q}^{\alpha} +  \hat{q}^{\alpha} \hat{\rho}  \hat{P}^{(-)\alpha}),  \label{eq:pollard}
\end{align}
where $\hat{M}^{+} = \sum_{\alpha} \hat{q}^{\alpha} \hat{P}^{(+)\alpha}$, $\hat{M}^{-} = \sum_{\alpha}  \hat{P}^{(-)\alpha}\hat{q}^{\alpha}$, $P^{\pm(\alpha)}_{ik} = q^{(\alpha)}_{ik} B^{\pm (\alpha)}_{ik}/2 \pi$. Since $\hat{M}$, $\hat{P}$, and $\hat{q}$ are all operators, not superoperators, they are of size $d\times d$, where $d$ is the dimension of the Hilbert space. This is in contrast with the superoperator $\hat{R}$, which is of size $d^2\times d^2$. Thus storing $\hat{M}$, $\hat{P}$, and $\hat{q}$ is feasible for the model pyrazine system, whereas storage of $\hat{R}$ is not. The secular Redfield tensor is also of interest, since it is used as a preconditioner in the numerical algorithm discussed in section \ref{sec:numerics}, and since it is used to compare to non-secular results below. In the secular approximation, the system density operator satisfies \cite{pauli1,pauli2}:
\begin{align}
& \frac{d}{dt} \rho_{ii} = \sum_{j\neq i} Z_{ij} \rho_{jj} - \rho_{ii} \sum_{j \neq i} Z_{ji}, \\
&\frac{d}{dt} \rho_{ij} = [-i\omega_{ij} - \gamma_{ij} ] \rho_{ij},
\end{align}
with $Z_{ij} = \sum_{\alpha }\Gamma^{(+)\alpha}_{ijji} + \Gamma^{(-)\alpha}_{ijji}$, and $\gamma_{ij} = \sum_{\alpha} (-\Gamma^{(+)\alpha}_{jjii} -\Gamma^{(-)\alpha}_{jjii} + \sum_{r} \Gamma^{(+)\alpha}_{irri} + \Gamma^{(-)\alpha}_{jrrj})$.
\subsection{Numerical results} \subsubsection{Reconstructed dynamics}The Hamiltonian (\ref{eq:hamiltonian}) was assembled in a direct product basis of eigenstates of the harmonic oscillator centered at $x=0$, using 25 basis functions for each vibrational mode. Diagonalization of the Hamiltonian yielded the 600 excited eigenstates of lowest energy, which were added to the 60 ground states of lowest energy for a combined Hilbert space of dimension $660$. The choice of 60 ground states was made to achieve convergence of the $S_1/S_2 \to S_0$ spontaneous emission rates of the 10 excited states of lowest energy. The choice of 600 excited states was made to achieve convergence of the $S_0 \to S_1/S_2$ excitation rates for the 60 ground states. We also confirmed that the steady state $S_1$ population did not change with the inclusion of more basis functions, $S_0$ eigenstates, or $S_1/S_2$ eigenstates. Unless otherwise indicated, the initial system density operator was taken to be Boltzmann distributed, such that $\hat{\rho}_0 = \mathrm{exp}[-\hat{H}_{\mathrm{S}}/k_{\mathrm{B}} T_{\mathrm{amb}}] / \mathrm{Tr}(\mathrm{exp} [-\hat{H}_{\mathrm{S}}/k_{\mathrm{B}} T_{\mathrm{amb}}])$, where $T_{\mathrm{amb}} =300$ K is the ambient temperature. These initial conditions describe a situation in which the system and bath are initially in thermal equilibrium. The master equation (\ref{eq:master_nonsec}) describes the subsequent dynamics induced by a radiation field that is suddenly turned on at $t=0$. Of interest is the adiabatic $S_1$ population, given by $\mathrm{Tr}[\hat{\rho}(t) \hat{P}^{(1)}_{\mathrm{ad}}] $, where the adiabatic projector for the $n$th electronic state is given by $\hat{P}^{(n)}_{\mathrm{ad}} = \ket{\tilde{\phi}_n} \bra{\tilde{\phi}_n}$, and $\ket{\tilde{\phi}_n}$ is the $n$th adiabatic Born-Oppenheimer state. The adiabatic populations are of interest since they describe non-adiabatic effects in the internal conversion process, whereas the diabatic populations are more closely related to optical spectra \cite{pyrazine_2s2m}. The $n$th adiabatic state is related to the diabatic electronic states through \cite{pyrazine_2s2m}
\begin{align}
\ket{\tilde{\phi}_n} = \sum_{m=1,2} \hat{Q}(\hat{x}_c,\hat{x}_t)_{nm} \ket{\phi_m},
\end{align}
where $\hat{Q}$ is the matrix that diagonalizes the potential energy part of $\hat{H}_{\mathrm{S}}$. The projector $\hat{P}^{(1)}_{\mathrm{ad}}$ can be written as an explicit function of $\hat{x}_c$ and $\hat{x}_t$, given by Eq. (6) in Ref. \cite{adiabatic_reference}. This expression was used to create the adiabatic projector in the eigenbasis of the position operators $\hat{x}_c$ and $\hat{x}_t$. The projector was then transformed into the harmonic oscillator energy eigenstate basis, and subsequently transformed into the eigenstate basis of the full Hamiltonian (\ref{eq:hamiltonian}).
\begin{figure}[t!]
	\centering
	\includegraphics[width=\columnwidth,trim={0cm 0cm 0cm 0cm},clip]{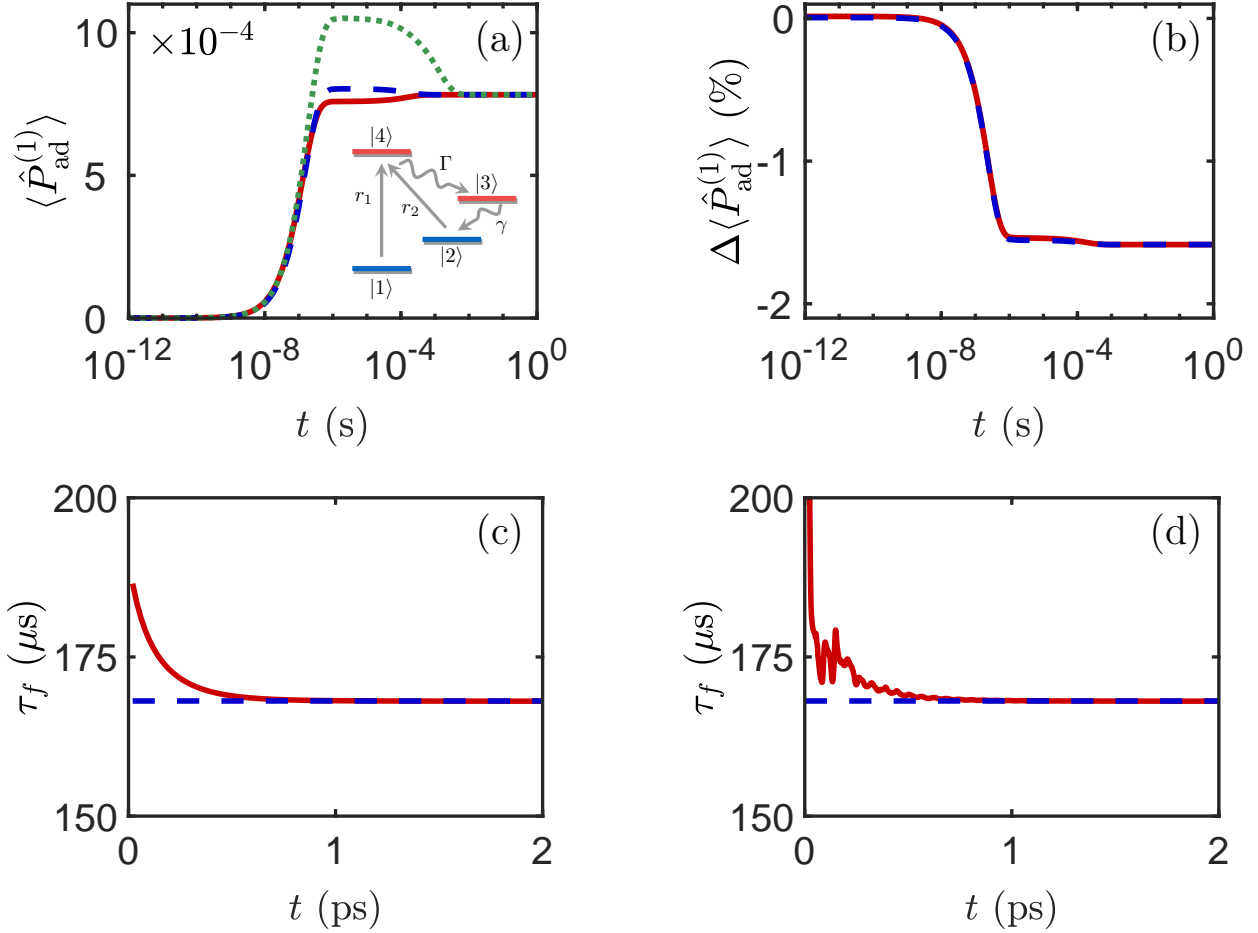}
	\caption{(a) Reconstructed dynamics of the adiabatic $S_1$ population in the secular approximation, using Boltzmann initial conditions (solid red) and $\hat{\rho}_0 = \ket{3} \bra{3}$ (dashed blue). Dotted green; the latter case with $\braket{1 \vert \hat{\mu} \vert i}$ reduced by a factor of $\sqrt{10}$ for all $i$. Inset: Energy diagram of the model four-level system used to describe the results of the main figure. Arrows and colors are as in Fig. \ref{fig:systems_schematic}. (b) Discussed in section \ref{subsec:modified_model} below: relative difference between populations with and without the secular approximation, $\Delta \braket{\hat{P}^{(1)}_{\mathrm{ad}}} = (\braket{\hat{P}^{(1)}_{\mathrm{ad}}}_{\mathrm{non-sec}}-\braket{\hat{P}^{(1)}_{\mathrm{ad}}}_{\mathrm{sec}})/\braket{\hat{P}^{(1)}_{\mathrm{ad}}}_{\mathrm{non-sec}}$, using a modified set of system parameters. (c) Inverse forward reaction rate $\tau_f$, calculated from solution of the secular master equation (solid red) and from the method described in this work (dashed blue). Regular system parameters and Boltzmann initial conditions are used. (d) As in (c), but without the secular approximation.}
	\label{fig:pyrazine_dynamics}
\end{figure}

The secular and non-secular dynamics of the adiabatic population were reconstructed using the method of section \ref{sec:numerics}. Three exponential functions were used in Eq. (\ref{eq:expansion}) both with and without the secular approximation. The three weight functions and three decay constants were obtained from five progress moments together with the condition $\sum_i f_i = \braket{\hat{\chi}(0)}$. The progress moments were obtained in less than a second in the secular approximation. The projection of the adiabatic population onto an exponential basis was performed by solving Eq. (\ref{eq:basis_solve}) using the procedure outlined in section \ref{subsection:projection}; Eq. (\ref{eq:basis_solve}) was solved in less than five seconds. The secular $S_1$ adiabatic dynamics are shown in Fig. \ref{fig:pyrazine_dynamics}(a) for the Boltzmann initial conditions (solid red), and, for comparison, for the initial conditions $\hat{\rho}_0 = \ket{3} \bra{3}$ (dashed blue). Here, state $\ket{3}$ is the energy eigenstate with the third lowest energy. The non-secular dynamics were found to be equivalent to the secular dynamics, as discussed below. For both sets of initial conditions, one observes three regimes. The first regime is associated with the smooth increase of the adiabatic $S_1$ population over several hundred nanoseconds. The second regime is associated with a dramatic change in the derivative of the adiabatic population, which nonetheless remains nonzero over a millisecond time scale. The third is the long-time steady-state regime in which the adiabatic population is constant. The value of the adiabatic population reached at the end of the first regime is hereafter referred to as its quasi-stationary value.
The long time stationary values in Fig. \ref{fig:pyrazine_dynamics}(a) are seen to be the same for both sets of initial conditions since the steady state is unique. However, different initial conditions result in different quasi-stationary values. In this case, for example, the quasi-stationary value obtained from Boltzmann initial conditions is seen to be lower than the stationary value, whereas it is higher than the stationary value for the initial conditions $\hat{\rho}_0 = \ket{3} \bra{3}$. Clearly it is advantageous to have been able to obtain such dynamics on the nanosecond and millisecond time scales without direct solution of the master equation. Moreover, this method allows for the calculation of both stationary and quasi-stationary expectation values of relevant operators. While the former quantity can be calculated by obtaining the unique stationary state $\hat{\rho}_s$, the latter quantities are more difficult to calculate. Indeed, if the quasi-stationary states are not unique, as is evident here, then it is not immediately clear which quasi-stationary state is reached for different initial conditions.
\subsubsection{Disparate time scales}To expose the origin of these three time scales, consider a simple rate-law description of the system in terms of four energy levels, where states $\ket{1}$ and $\ket{2}$ are vibrational states of $S_0$, state $\ket{4}$ represents an $S_2$ level, and state $\ket{3}$ represents an $S_1$ level, the ``product'' [inset of Fig. \ref{fig:pyrazine_dynamics}(a)]. Excitation out of state $\ket{1}$ occurs at rate $r_1$, which is in general different from the excitation rate $r_2$ out of state $\ket{2}$. Non-radiative decay from state $\ket{4}$ occurs at a rate $\Gamma \gg r_1, \ r_2, \ \gamma$, where $\gamma$ is the spontaneous emission rate between states $\ket{3}$ and $\ket{2}$. In the full model of pyrazine, excitation from the Boltzmann initial state occurs predominantly from the ground state. However, spontaneous emission between the lowest-lying excited state ($\ket{3}$ in this simple model) transfers population to both of the two $S_0$ eigenstates of lowest energy ($\ket{1}$ and $\ket{2}$ in this model). Hence excitation and spontaneous emission are not associated with one $S_0$ state only. To understand the basic physics of this situation, the spontaneous emission rate between states $\ket{3}$ and $\ket{1}$ is set to zero in this four-level model. 

In the limit $\Gamma \gg \gamma, r_1, r_2$, excitation from $\ket{1}$ and $\ket{2}$ to states $\ket{4}$ may be replaced with direct excitation to state $\ket{3}$, yielding the rate equations
\begin{subequations} \label{eq:rate_law_pyrazine}
	\begin{align}
	& \dot{\rho}_{11}(t) = -r_1 \rho_{11}(t)  \\
	&  \dot{\rho}_{22}(t) = -r_2 \rho_{22}(t) + \gamma \rho_{33}(t)  \\
	& \dot{\rho}_{33}(t) =r_1 \rho_{11}(t) +  r_2 \rho_{22}(t) - \gamma \rho_{33}(t).
	\end{align}
\end{subequations}
To see the origin of the quasi-stationary regime, consider $\rho_{33}(t)$. Solving Eq. (\ref{eq:rate_law_pyrazine}) analytically and invoking the fact that $r_1, \ r_2 \ll \gamma$ yield
\begin{align}
\label{eq:rate_law_solve}
\rho_{33}(t) = \frac{r_2}{\gamma} \bigg( 1 - \frac{r_1}{r_2} e^{-\gamma t} + \left( \frac{r_1}{r_2}-1 \right) e^{-r_1 t} \bigg).
\end{align}
From Eq. (\ref{eq:rate_law_solve}) it is clear that for $r_1 = r_2$, the system equilibrates on the spontaneous emission time scale $\gamma^{-1}$. However, for $r_1 \neq r_2$, there is a fast time scale $\gamma^{-1}$ associated with equilibration within the $\ket{2}, \ \ket{3}$ manifold, and a long time scale $r_1^{-1}$ associated with excitation from state $\ket{1}$. Since the initial dynamics occur on a time scale of $\gamma^{-1}$, and the subsequent dynamics occur on a much longer time scale of $r^{-1}$, the value of $\rho_{33}(t)$ for $\gamma^{-1} \ll t \ll r^{-1}$ is interpreted as its quasi-stationary value. At these times, $\mathrm{exp}(-\gamma t) \approx 0$ and $\mathrm{exp}(-r_1 t) \approx 1$, so the quasi-stationary value is obtained from Eq. (\ref{eq:rate_law_solve}) as $\rho_{33} =r_1/\gamma$. The stationary value is obtained from Eq. (\ref{eq:rate_law_solve}) in the limit $t \to \infty$ as $\rho_{33} =r_2 / \gamma$. An increase from the quasi-stationary to stationary value occurs if $r_2 > r_1$, and a decrease occurs if $r_2 < r_1$. These two time scales are evident in the pyrazine dynamics shown in Fig. \ref{fig:pyrazine_dynamics}(a). If the filtering of incident light were included \cite{hoki,timur2}, the time scale $r^{-1}$ would be even longer. To emphasize the dependence of the quasi-stationary behavior on the relative difference between excitation rates, we have also artificially reduced $\braket{1 \vert \hat{\mu} \vert i}$ and $\braket{i \vert \hat{\mu} \vert 1}$ by a factor of $\sqrt{10}$ for all $i$, leading to a 10-fold reduction in the excitation rate out of the ground state. The adiabatic population for this case, with the initial conditions $\hat{\rho}_0 = \ket{3}\bra{3}$, is shown with a dotted green line in Fig. \ref{fig:pyrazine_dynamics}(a). The difference between the quasi-stationary and stationary values is seen to increase, as expected. 

Also of interest is the $S_0 \to S_1$ reaction rate. Two different time scales are of interest. The first is for times before $t^*$, where the dynamics are not expected to follow a rate law. The second time scale is that of $t>t^*$, at which time the dynamics are expected to follow a rate law. We define the phenomenological reaction rate $k_f$ such that $k_f \approx \partial_t \braket{\hat{P}_{\mathrm{ad}}^{(1}} $ for $t> t^*$.
By defining a time-dependent forward rate $k_f(t) =\partial_t \braket{\hat{P}^{(1)}_{\mathrm{ad}}(t)} $, calculating $k_f(t)$ through direct simulation, and plotting the results, the phenomenological forward rate is obtained graphically as the plateau value of $k_f(t)$. The time at which this plateau is reached defines the time $t^*$, which, it should be noted, is not known \textit{a priori}.

The inverse forward rate $\tau_f \equiv k_f^{-1}$, calculated from progress moments using Boltzmann initial conditions and the secular approximation, is shown as a dashed blue line in Fig. \ref{fig:pyrazine_dynamics}(c), and is compared to the full  simulation results shown with a solid red line. The analogous non-secular inverse result is plotted in Fig. \ref{fig:pyrazine_dynamics}(d). Excellent agreement between simulated and calculated values of $\tau$ by the plateau time $t^*$ are shown in Fig. \ref{fig:pyrazine_dynamics}(c) and (d). Because of the disparity in dynamical time scales discussed above, a single Laplace transform was used to improve the accuracy of the fast decay rate of the progress variable. This additional Laplace transform, described in Appendix \ref{app:laplace}, yielded only a small change in the fast decay of the progress variable under Boltzmann initial conditions. However, for the initial conditions $\hat{\rho}_0 = \ket{3} \bra{3}$ the fast decay rate was corrected by 19\% (not shown). 

The forward reaction rate is seen to stabilize after approximately 1 ps, which defines the time $t^*$ \cite{2001,kapral}. This stabilization time is determined by the time taken for bath-induced decay from the bright eigenstates to the minimum of the $S_1$ potential well. When the system-bath coupling is enhanced or diminished by a  factor $\lambda$, the stabilization time is found to scale linearly with $\lambda^{-1}$ (not shown). When propagating the secular master equations this variation is irrelevant, since changing the system-bath coupling only changes the natural time unit of the system. Since both $t^*$ and the minimum step size $\Delta t$ increase, the total number of time-steps is fixed. In the non-secular case, however, the natural oscillation frequencies of the system remain fixed even as the system-bath coupling is reduced, so that $\Delta t$ is also fixed. Since $t^*$ increases while $\Delta t$ remains fixed, the total number of time-steps increases. Therefore, the advantage of using the method presented here becomes particularly clear when considering non-secular dynamics. 
\subsubsection{Non-secular effects} Non-secular effects are observed in the transient behavior of the reaction rate in Fig. \ref{fig:pyrazine_dynamics}(d). In particular, the oscillations present in the reaction rate are not present in the secular case, Fig. \ref{fig:pyrazine_dynamics}(c). However, by $t^*$ the rates are seen to reach the same value, implying that the effect of coherence is transient and irrelevant on the gross time scales of Fig. \ref{fig:pyrazine_dynamics}(a). Hence, a further advantage of this approach is the opportunity to judge the validity of the secular approximation without requiring  simulation to $t^*$. 

To understand why the plateau values of $k_f(t)$ are the same in both the secular and non-secular cases [Fig. \ref{fig:pyrazine_dynamics}(c) and (d)], consider that bath-induced transitions favor downhill energy transfer at finite temperature, and so by the plateau time $t^*$, the bath will have transferred the majority of the excited population to lower energy $S_1/S_2$ states. In the absence of solar coupling, the effect of the bath would be to redistribute the excited population into an equilibrium state with respect to the bath. Under Redfield dynamics, this equilibrium state is a Boltzmann distribution with respect to temperature $T = 300$ K \cite{agarwal_steady}. Since the rates of excitation and spontaneous emission are small compared to bath-induced relaxation, the effect of the bath is indeed to transfer the majority of the excited population into such a state of instantaneous equilibrium. Since a Boltzmann distribution is devoid of coherences in the energy eigenbasis, the density operator is diagonal in this subspace. Therefore, at times $t>t^*$, the secular and non-secular values of $\braket{\hat{P}^{(1)}_{\mathrm{ad}}}$, which are determined by the density operator in the $S_1/S_2$ subspace, are identical. For this reason, the $S_0 \to S_1$ reaction rate is independent of non-secular effects. 

Indeed, for weak excitation the reaction rate is expected to equal the excitation rate multiplied by the reaction yield, as described in \ref{subsection:example_systems_chemical}. The need to calculate the reaction rate lies in determining the reaction yield. However, a straightforward calculation shows that only a few of the lowest lying $S_1/S_2$ energy eigenstates are populated in a Boltzmann distribution at $T= 300$ K, and that each of these states is found to be almost exclusively $S_1$. Therefore, all of the population that is excited from $S_0$ is converted to $S_1$, and so the $S_0 \to S_1$ reaction rate $k_f$ is simply the rate of excitation. 

To see when non-secular effects could be important, consider that population is transferred from $S_1$ to $S_0$ through spontaneous emission. The rate at which $S_1 \to S_0$ population transfer occurs depends upon the nature of the $S_1$ states that are occupied. For example, an in-phase superposition of low-lying $S_1$ energy eigenstates (admittedly rare under thermal conditions) may be associated with a higher rate of population transfer through spontaneous emission than a statistical mixture of these states \cite{dicke}.  That is,  non-secular effects can appear in the rate of $S_1 \to S_0$ population transfer if the nature of the $S_1$ state is changed. \subsubsection{A modified model}
\label{subsec:modified_model}
To examine non-secular effects we consider an arbitrarily modified 2-mode pyrazine model. Specifically, we make three modifications to the model to change the Boltzmann distribution, which is devoid of coherences. First, consider that changing the temperature of the bath that couples to $\hat{x}_t$ can disrupt the Boltzmann distribution, since there is no longer a single temperature with respect to which a Boltzmann distribution would equilibrate. In this case, coherences can form in the excited manifold. To examine this, we set the temperature of the bath that couples to $\hat{x}_t$ to an extreme value of 1200 K. Second, we reduce the energy splitting between the three lowest-lying $S_1/S_2$ eigenstates to $10^{-7}$ eV, so as to enhance the magnitude of any generated coherence. Third, we note that the effect of coherence on the collective spontaneous emission rate is closely related to the alignment of transition dipole matrix elements \cite{timur_prl}. We therefore replace all transition dipole matrix elements with their absolute values. In this way, all incoherent transitions induced by solar coupling to the $S_0-S_1$  or $S_0-S_2$ dipole operator occur through aligned dipole matrix elements. 

The resultant difference between secular and non-secular values of $\braket{\hat{P}^{(1)}_{\mathrm{ad}}}$ is show in Fig. \ref{fig:pyrazine_dynamics}(b). The red line corresponds to Boltzmann initial conditions, and the dashed blue line to the initial conditions $\hat{\rho}_0 = \ket{3} \bra{3}$. In both cases the relative change is still small, 1.5\% in steady state. To understand why this is the case, we also calculated the steady state $\hat{\rho}_s$ with spontaneous emission artificially excluded, so as to examine the nature of the instantaneous equilibrium state formed in the $S_1/S_2$ manifold. The results (not shown) indicate that, even with the modified parameters, the magnitude of the coherence in the instantaneous equilibrium state is small compared to the populations.

While this example is certainly artificial, and only yields a small effect, its key feature  is that a Boltzmann distribution is not formed.  This requirement is fulfilled in a variety of physical systems, even in systems where there is only solar coupling and system-bath coupling at $T= 300$ K. In the case of retinal isomerization, for example, there are stable potential wells associated with both the \textit{cis} and \textit{trans} conformations, both of which are accessible from the excited manifold \cite{timur1}. In this case, the  bath-induced decay does not lead to a Boltzmann distribution. As we will show in future work, the effect of coherence in minimal models of retinal can be substantial \cite{second_paper}. The time $t^*$ will also be shown to be quite long (approximately 80 ps), which further necessitates the use of the method presented here. Note also that if the $S_1$ and $S_0$ states in model pyrazine were coupled through the bath, a Boltzmann distribution would not be formed in this system, either.

The advantage of the approach introduced here is clear from the example examined above. In this case, the non-secular simulation to $t^* = 2$ ps using a 0.5 fs time-step took two hours of central processing unit (CPU) time, while the calculation of the steady state $\hat{\rho}_s$ and the progress moments took 1-3 minutes each. The resultant dynamics over a millisecond time scale serve as a clear demonstration of the power of this method, suggesting application in the simulation of multi-scale dynamics. The computational time increases as non-secular effects become  significant; however, as we will show in a future publication \cite{second_paper}, the computational time is still reasonable when non-secular effects are more significant. For example, in a two-state, two-mode model describing \textit{cis}-\textit{trans} isomerization of retinal \cite{hahn1,balzer,timur1,timur2}, the system is characterized by over 700 eigenstates and significant population-coherence coupling. However, the calculation of $\hat{\rho}_s$ and the $I_n$ takes only 30-60 minutes each, while a direct simulation to $t^*$ takes several days.

The approach introduced in this work can provide an efficient alternative to direct solution of a quantum master equation, as demonstrated for model pyrazine above. We remark, however, that in both cases it is necessary to calculate the effect of $\hat{\mathcal{L}}$, which is an order $d^2$ operation within the secular approximation, order $d^3$ under Redfield dynamics, and order $d^4$ in general. Thus our method is limited by the same adverse scaling as the quantum master equation. However, if one is interested in only a few degrees of freedom of the system,  the system can be chosen to have a fairly small dimension. Here we have modeled the bath with a Redfield master equation, but a detailed description of the inactive modes of the molecule can be included with such methods such as the HEOM \cite{heom_pyrazine} or the multi-layer multi-configuration time-dependent Hartree method \cite{multilayer_MCTDH}. Each of these methods can be used to calculate the memory kernel \cite{jesenko,wilner_1,wilner_2} and hence the Liouville superoperator (see Appendix). Combining these well-established methods with our approach to calculate the light-induced dynamics of large molecules is an area of interest for future work.
\section{Conclusion}
\label{sec:conclusion}
We have introduced a new approach to reconstructing the dynamics of molecular processes in open systems, i.e. those coupled to an environment. This technique is particularly useful for quantum mechanical systems subject to weak, incoherent excitation, such as sunlight or noise. Such systems are ubiquitous in nature, and of particular interest in biological and chemical physics. We have provided a computational algorithm for this technique and presented three examples. The technique was shown to be accurate when applied to a model pyrazine system characterized by over 600 eigenstates and significant population-coherence coupling. An efficient algorithm for calculating the non-equilibrium steady state under Redfield dynamics has also been examined, and we have shown how quasi-stationary values of reaction observables can be calculated without propagating a differential equation. We expect that this approach will find application in a variety of systems subject to weak, incoherent excitation.
\section{Acknowledgments}
\label{sec:acknowledgements}
We thank Cyrille Lavigne for enlightening discussions. This work was supported by the U.S. Air Force Office of Scientific Research (AFOSR) under contract number FA9550-17-1-0310.
{\parindent0pt \section*{Appendix} }
\renewcommand{\theequation}{\Alph{subsection}\arabic{equation}}
\setcounter{equation}{0}
\renewcommand{\thesubsection}{\Alph{subsection}}
\setcounter{section}{0}
\subsection{Generalization to non-Markovian dynamics}
\label{app:non_markov}
To prove Eq. (\ref{eq:extension}) of the main text, start with the master equation $\partial_t \hat{\rho}(t) = \int_{0}^{t} d\tau \ \hat{\mathcal{K}}(t-\tau)  \hat{\rho}(\tau)$, valid for non-Markovian systems with no initial correlations, to write \cite{non_markov_rates}
\begin{align}
& \hat{0} = \int_{0}^{\infty} dt \ t^n \ \bigg(- \partial_t [ \delta \hat{\rho}(t)] + \int_{0}^{t}d\tau \ \hat{\mathcal{K}}(t-\tau)  \hat{\rho}(\tau) \bigg), \label{eq:identity}
\end{align}
where $\delta \hat{\rho}(t) \equiv \hat{\rho}(t) - \hat{\rho}_s$, and we have used the fact that $\hat{\rho}_s$ is not a function of time to write, $\partial_t [\hat{\rho}(t)] = \partial_t [\delta \hat{\rho}(t)]$. Writing $\hat{\rho}(\tau)  = [\hat{\rho}(\tau) - \hat{\rho}_s]  + \hat{\rho}_s$ yields
\begin{align}
& \hat{0} = \int_{0}^{\infty} dt \ t^n  \ \bigg( - \partial_t [\delta \hat{\rho}(t)]  +  \int_{0}^{t} d\tau \ \hat{\mathcal{K}}(t-\tau) \hspace*{0.1cm} [ \delta \hat{\rho}(\tau)  + \hat{\rho}_s] \bigg) \nonumber \\
& = \hat{S}_n + \hat{T}_n + \int_{0}^{\infty} dt \ t^n \int_{0}^{t} d\tau \ \hat{\mathcal{K}}(t-\tau) \hspace*{0.1cm}  \delta \hat{\rho}(\tau)  , \label{eq:identity2}
\end{align}
where the known quantities $\hat{S}_n$ and $\hat{T}_n$ are
\begin{align}
& \hat{S}_n =  - \int_{0}^{\infty} dt \ t^n \ \partial_t [\delta \hat{\rho}(t)] = \begin{dcases}
\hat{\rho}_0 - \hat{\rho}_s, & \text{if } n= 0 \\
 n \cdot\delta \hat{\rho}_{n-1},  &\text{if } n \neq 0, \label{eq:bn}
\end{dcases} \\
& \hat{T}_n =\int_{0}^{\infty} dt \ t^n \bigg(\int_{0}^{t} d\tau \ \hat{\mathcal{K}}(t-\tau) \hat{\rho}_s \bigg). \label{eq:T_def}
\end{align}
Equation (\ref{eq:bn}) follows from integration by parts, using the fact that $\delta \hat{\rho}(\infty) = \hat{0}$, $\delta \hat{\rho}(0)= \hat{\rho}_0 - \hat{\rho}_s$, and $t^n\vert_{t=0} = 0$ for $n>0$. Since $\int_{0}^{t} d\tau \ \hat{\mathcal{K}}(\tau) \hat{\rho}_s \to \hat{0}$ as $t \to \infty$, the time integral of the bracketed term in Eq. (\ref{eq:T_def}) is finite.  

To evaluate the integral in Eq. (\ref{eq:identity2}) write $t^n$ in terms of the variables $(t-\tau)$ and $\tau$, through $t^n= [(t-\tau) + \tau]^n = \sum_{k=0}^{n} \binom{n}{k} (t-\tau)^{n-k} \tau^{k}$, where $\binom{n}{k} = n!/[k! (n-k)!] $ is the binomial coefficient, with the result
\begin{align}
& \hat{0} = \hat{S}_n +\hat{T}_n + \sum_{k=0}^{n} \binom{n}{k} \int_{0}^{\infty} dt \ \int_{0}^{t} d\tau \  (t-\tau)^{n-k} \ \hat{\mathcal{K}}(t-\tau) \hspace*{0.15cm} \delta \hat{\rho}(\tau) \tau^k  \nonumber \\
& =\hat{S}_n +\hat{T}_n  + \sum_{k=0}^{n} \binom{n}{k} \bigg( \int_{0}^{\infty} dt_1 \   (t_1)^{n-k}  \ \hat{\mathcal{K}}(t_1) \bigg) \bigg( \int_{0}^{\infty} dt_2 \ t_2^k \ \delta \hat{\rho}(t_2) \bigg) \nonumber \\
& = \hat{S}_n +\hat{T}_n + \sum_{k=0}^{n} \binom{n}{k} \hat{\mathcal{L}}_{n-k} [\delta \hat{\rho}_k] \label{eq:start},
\end{align}
where in the last line we have defined $\hat{\mathcal{L}}_n = \int_{0}^{\infty} dt \ t^n \hat{\mathcal{K}}(t)$. This result was obtained using the Laplace transform convolution theorem:
\begin{align}
& \lim_{\alpha \to 0 } \int_{0}^{\infty} dt \ e^{-\alpha t} \ \bigg( \int_{0}^{t} d\tau \ f(t-\tau) g(\tau) \bigg) = \lim_{\alpha \to 0} \bigg( \int_{0}^{\infty} dt_1 \ e^{-\alpha t_1} f(t_1) \bigg) \cdot \bigg( \int_{0}^{\infty} dt_2 \ e^{-\alpha t_2} g(t_2) \bigg) \nonumber \\
& = \bigg( \int_{0}^{\infty} dt_1 \ f(t_1) \bigg) \cdot \bigg( \int_{0}^{\infty} dt_2 \  g(t_2) \bigg). \label{eq:laplace_theorem}
\end{align}
Here $f(t)$ and $g(t)$ are test functions, and the last line follows if $\int_{0}^{\infty} dt \ f(t) $ and $\int_{0}^{\infty} dt \ g(t) $ exist. The evaluation of the integral in Eq. (\ref{eq:start}) then follows by writing the integrand in terms of its matrix elements,
\begin{align}
 & \int_{0}^{\infty} dt \ \int_{0}^{t} d\tau \  (t-\tau)^{n-k} \ \bra{i} \hat{\mathcal{K}}(t-\tau) \hspace*{0.15cm} \delta \hat{\rho}(\tau) \ket{j} \tau^k  = \sum_{lm}  \int_{0}^{\infty} dt \ \int_{0}^{t} d\tau \ (t-\tau)^{n-k} \times \nonumber \\
 & \mathcal{K}_{ijlm}(t-\tau) \delta \rho_{lm}(\tau)\tau^k \equiv \sum_{lm} \int_{0}^{\infty} dt \ \int_{0}^{t} d\tau \ f^{(k)}_{ijlm} (t-\tau) \hspace*{0.1cm} g^{(k)}_{lm}(\tau),
\end{align}
which is the form of Eq. (\ref{eq:laplace_theorem}).  

Consider now that $\delta \hat{\rho}_n$ enters into Eq. (\ref{eq:start}) under the action of $\mathcal{\hat{L}}_0$ through $\hat{\mathcal{L}}_0[\delta \hat{\rho}_n]$. Since the steady-state solution satisfies $\partial_t \hat{\rho}_s = \lim_{t \to \infty} \int_{0}^{t} d\tau \ \hat{\mathcal{K}}(t-\tau) \hat{\rho}(\tau) = \hat{\mathcal{L}}_0 \hat{\rho}_s = \hat{0}$, there exists a non-trivial solution to the equation $\hat{\mathcal{L}}_0 \hat{\rho}_s = \hat{0}$, and so $\hat{\mathcal{L}}_0$ is singular. Therefore, $\delta\hat{\rho}_n$ cannot be isolated in Eq. (\ref{eq:start}) by inverting $\hat{\mathcal{L}}_0$. Hence, another property of $\delta \hat{\rho}_n$ must be specified so that it can be uniquely determined. One such property is that each operator $\delta \hat{\rho}_n$ is traceless when population is conserved, since $\mathrm{Tr}[\hat{\rho}(t)] = \mathrm{Tr}[\hat{\rho}_s] = 1$. Therefore, we add the term $w \hat{\mathcal{T}} [{\delta \hat{\rho}_n}]$, where the action of $\hat{\mathcal{T}}$ on $\delta \hat{\rho}_n$ is given by $w\hat{\mathcal{T}} [\delta \hat{\rho}_n] = w \mathrm{Tr}[\delta \hat{\rho}_n] \ket{1}\bra{1}$, as discussed in the main text. Adding this term to the right-hand side of Eq. (\ref{eq:start}) yields
\begin{align}
& - \hat{\mathcal{L}}_0[\delta \hat{\rho}_n] - w \hat{\mathcal{T}}[\delta \hat{\rho}_n]=  \hat{S}_n + \hat{T}_n + \sum_{k=0}^{n-1} \binom{n}{k} \hat{\mathcal{L}}_{n-k}[\delta \hat{\rho}_k].  \label{eq:intermediate}
\end{align}
The sum on the right-hand side of Eq. (\ref{eq:intermediate}) is understood to be zero for $n=0$. Below we show explicitly that Eq. (\ref{eq:intermediate}) implies both Eq. (\ref{eq:start}) and $\mathrm{Tr}[\delta \hat{\rho}_n] = 0$, and that $[\hat{\mathcal{L}}_0 +w \hat{\mathcal{T}}]^{-1}$ exists. The operator $\delta \hat{\rho}_n$ can now be isolated, through
\begin{align}
\delta \hat{\rho}_n = - [ \hat{\mathcal{L}}_0 + w \hat{\mathcal{T}} ]^{-1 } \begin{dcases}
\hat{\rho}_0 - \hat{\rho}_s +\hat{T}_0  ,  \vphantom{\frac{0}{0}} & \text{if } n= 0 \\
n \cdot \delta \hat{\rho}_{n-1} + \hat{T}_n + \sum_{k=0}^{n-1} \binom{n}{k} \hat{\mathcal{L}}_{n-k} [\delta \hat{\rho}_k],  &\text{if } n \neq 0,
\end{dcases} \label{eq:final_delta_rho}
\end{align}
which is Eq. (\ref{eq:extension}) in the main text. In the Markovian case, $\hat{\mathcal{K}}(t) = \hat{\mathcal{L}}_0 \cdot  \delta(t)$, we have $\hat{\mathcal{L}}_{m>0}  = \int_{0}^{\infty} dt  \ [\delta(t) \hspace*{0.05cm} t^m  ] \hspace*{0.1cm} \hat{\mathcal{L}}_0  = \hat{0}$ and $\hat{T}_n = \int_{0}^{\infty} dt \ t^n \hspace*{0.1cm} [\hat{\mathcal{L}}_0 \hat{\rho}_s]  = \hat{0}$. Hence non-Markovian corrections to $\delta \hat{\rho}_n$ arise from $\hat{T}_n$ and from the sum over $\hat{\mathcal{L}}_{n-k} [\delta \hat{\rho}_k]$, where $n-k$ is positive.

Three notes are in order:
\begin{enumerate}
	\item To see that the inclusion of $w \hat{\mathcal{T}}$ in Eq. (\ref{eq:intermediate}) yields both Eq. (\ref{eq:start}) and  $\mathrm{Tr}[\delta \hat{\rho}_n] = 0$, consider that $\mathrm{Tr}[\hat{\mathcal{K}}(t) \hat{z}] = 0$ for any operator $\hat{z}$, which is proved below.
	Then applying the trace to Eq. (\ref{eq:intermediate})
	and noting that $\mathrm{Tr}[\hat{S}_n]  = 0$ yield $w \mathrm{Tr}(\ket{1}\bra{1})\mathrm{Tr}[ \delta \hat{\rho}_n] = 0$, and thus $\mathrm{Tr}[ \delta \hat{\rho}_n] = 0$. Substituting this result back into Eq. (\ref{eq:intermediate}) then recovers Eq. (\ref{eq:start}). Hence Eq. (\ref{eq:intermediate}) implies both Eq. (\ref{eq:start}) and $\mathrm{Tr}[ \delta \hat{\rho}_n] = 0$.
	\item  Adding the superoperator $w \hat{\mathcal{T}}$ yields an invertible superoperator $\hat{\mathcal{L}}_0 + w \hat{\mathcal{T}}$ in Eq. (\ref{eq:intermediate}), since the only solution to the equation
	\begin{align}
	[\hat{\mathcal{L}}_{0}+ w \hat{\mathcal{T}}] \hat{z} = \hat{0} \label{eq:trivialB}
	\end{align}
	is the trivial solution $\hat{z}= \hat{0}$. To see why this is the case, consider the trace over each side. The property $\mathrm{Tr}[\hat{\mathcal{L}}_0 \hat{z}] = \int_{0}^{\infty} dt \ \mathrm{Tr}[\hat{\mathcal{K}}(t) \hat{z}]  = 0$ for arbitrary $\hat{z}$ then implies that $\mathrm{Tr}[\hat{z}] = 0$. Substituting this result into Eq. (\ref{eq:trivialB}) yields $\hat{\mathcal{L}}_0 \hat{z} = \hat{0}$. If the steady state is unique, then the only solution to this equation is $\hat{z} = \beta \hat{\rho}_s$ for some $\beta$, since $\hat{\mathcal{L}}_0 \hat{\rho}_s  = \hat{0}$. But since $\mathrm{Tr}[\hat{\rho}_s] = 1$ and $\mathrm{Tr}[\hat{z}]  =0$, the only possible value of $\beta$ is $\beta = 0$. Hence the only solution is the trivial solution, and so $[\hat{\mathcal{L}}_{0}+ w \hat{\mathcal{T}}]^{-1}$ exists. \\ \\
	To see why $\mathrm{Tr}[\hat{\mathcal{K}}(t) \hat{z}]  = 0$, we
	use the  exact form of the memory kernel derived from Nakajima-Zwanzig formalism. In the absence of initial system-bath correlations, the action of the memory kernel in the Schr\"odinger picture is given by \cite{breuer}
	\begin{align}
	& \hat{\mathcal{K}}(t- \tau) \hat{\rho}_{\mathrm{S}}(\tau) = - \frac{i}{\hbar} \mathrm{Tr}_{\mathrm{B}} \bigg( [\hat{H}_{\mathrm{T}}, \hat{\rho}_{\mathrm{S}}(\tau) \otimes \hat{\rho}_{\mathrm{B}}] \bigg) \delta(t-\tau)  \nonumber \\
	&- \frac{1}{\hbar^2}  \mathrm{Tr}_{\mathrm{B}} \bigg( [\hat{H}_{\mathrm{T}}, \hat{\mathcal{G}}(t-\tau) \hat{\mathcal{Q}} [\hat{H}_{\mathrm{T}}, \hat{\rho}_{\mathrm{S}}(\tau) \otimes \hat{\rho}_{\mathrm{B}}]] \bigg) , \label{eq:nakajima}
	\end{align}
	where $\hat{\rho}_{\mathrm{S}}(t) = \mathrm{Tr}_{\mathrm{B}}[\hat{\rho}_{\mathrm{T}}(t)]$ is the time-dependent system density operator and $\hat{\rho}_{\mathrm{B}} = \mathrm{Tr}_{\mathrm{S}} [ \hat{\rho}_{\mathrm{T}}(0)]$ is the initial bath density operator ($\hat{\rho}_{\mathrm{T}}$ is the total system-plus-bath density operator, while $\mathrm{Tr}_{\mathrm{S}}$ and $\mathrm{Tr}_{\mathrm{B}}$ denote a trace over the system and bath, respectively). The total system-plus-bath Hamiltonian is $\hat{H}_{\mathrm{T}}$, while $\hat{\mathcal{G}}(t)$ is a propagator and $\hat{\mathcal{Q}}$ is a projection superoperator ($\hat{\mathcal{G}}(t)= \mathrm{exp}(- i \hat{\mathcal{Q}} \hat{\ell}   t /\hbar)$, where $\hat{l} \hat{v}= [\hat{H}_{\mathrm{T}}, \hat{v}]$, $\hat{\mathcal{Q}} \hat{v} = \hat{v} - \mathrm{Tr}_{\mathrm{B}}(\hat{v}) \otimes \hat{\rho}_{\mathrm{B}}$, and $\hat{v}$ is an arbitrary operator over the combined system and bath). Applying an arbitrary system operator $\hat{z}$ and tracing over the system in Eq. (\ref{eq:nakajima}) yield
	\begin{align}
	\mathrm{Tr}_{\mathrm{S}}[\hat{\mathcal{K}}(t) \hat{z}] = - \frac{i}{\hbar} \mathrm{Tr}_{\mathrm{SB}} \bigg( [\hat{H}_{\mathrm{T}}, \hat{z} \otimes \hat{\rho}_{\mathrm{B}}]   \bigg) \delta(t) - \frac{1}{\hbar^2}  \mathrm{Tr}_{\mathrm{SB}} \bigg( [\hat{H}_{\mathrm{T}}, \hat{\mathcal{G}}(t) \hat{\mathcal{Q}} [\hat{H}_{\mathrm{T}}, \hat{z} \otimes \hat{\rho}_{\mathrm{B}}]] \bigg) = 0,
	\end{align}
	where $\mathrm{Tr}_{\mathrm{SB}}$ denotes a trace over the combined system and bath, and in the last line we have used the cyclic property of the trace. Thus $\mathrm{Tr}_{\mathrm{S}} [\hat{\mathcal{K}}(t) \hat{z}] = 0$ for arbitrary $\hat{z}$. Note that the time-independent Liouville superoperator $\hat{\mathcal{L}}$ considered in the main text is typically obtained from the Markovian contribution to the dynamics, through $\hat{\mathcal{L}} = \hat{\mathcal{L}}_0$.
	\item Consider a situation in which population is not conserved and the only steady state is $\hat{\rho}_s = \hat{0}$. This case applies, for example, to energy transfer from a photosynthetic system to a sink after initial excitation. Then Eq. (\ref{eq:intermediate}) can be re-derived for $\hat{\mathcal{L}}_0 [\int_{0}^{\infty} dt \ t^n \hspace*{0.1cm} \hat{\rho}(t)]$ in place of $\hat{\mathcal{L}}_0 [\delta \hat{\rho}_n]$. In this case $w \hat{\mathcal{T}}$ is not included since $\mathrm{Tr}[\delta \hat{\rho}_n]  \neq 0$.
	Moreover, in this case $\hat{T}_n = \hat{0}$ since $\hat{\rho}_s = \hat{0}$. Finally, since the only solution to $\hat{\mathcal{L}}_0 \hat{z} = \hat{0} $ is the null operator $\hat{z} = \hat{0}$, the superoperator $\hat{\mathcal{L}}_0 $ is invertible, and so $\int_{0}^{\infty} dt \ \hat{\rho}(t) = -\hat{\mathcal{L}}_0^{-1} \hat{\rho}_0$. For photosynthetic light harvesting systems the average time spent on the $m$th site, $\int_{0}^{\infty} dt \ \bra{m} \hat{\rho}(t) \ket{m}$, determines the transfer efficiency. The efficiency is thus independent of non-Markovian effects, which was proved in a different way in Ref. \cite{jesenko}. However, alternative definitions of the efficiency that are determined by $\int_{0}^{\infty} dt \ t^n \bra{m} \hat{\rho}(t) \ket{m}$ for $n \neq 0$ do depend upon non-Markovian effects in general, which is evidenced by the sum over $\hat{\mathcal{L}}_{n-k}[\delta \hat{\rho}_k]$ in Eq. (\ref{eq:intermediate}). It is interesting to consider that, in general, the zeroth progress moment depends upon non-Markovian effects through $\hat{T}_0$, and that non-Markovian effects only disappear when $\hat{\rho}_s = \hat{0}$.
\end{enumerate}
\subsection{Progress moments from the hierarchical equations of motion}
\label{app:heom}
\setcounter{equation}{0}
Here we show how progress moments can be obtained from the hierarchical equations of motion (HEOM). The HEOM provide a method of calculating the evolution of the density operator in the non-perturbative and non-Markovian regime \cite{kubo_heom_1989,tanimura1990,better_markov_heom1,ishizaki2009unified} using a time-independent Liouville superoperator. Application of the HEOM does not require the use of a memory kernel, although the memory kernel can be obtained through propagation of the HEOM if desired, as described in Ref. \cite{jesenko}. In the HEOM formalism, a set of auxiliary density operators $\hat{\rho}_{\mathbf{n}}(t)$ are introduced, where $\mathbf{n}$ is a vector that labels each density operator, and $\hat{\rho}_{\mathbf{n} =\mathbf{ 0}}(t)$ is the system density operator. We follow Refs. \cite{heom_time_independent} and \cite{heom_big} and define a linear vector space that contains the set of all auxiliary density matrices. In particular, we define the set of all auxiliary density matrices as $\newket{\boldsymbol{\rho}}= \{ \lket{\rho_{\mathbf{n} =\mathbf{0}}}, \lket{\rho_{\mathbf{n} \neq \mathbf{0}  }} \}$, where $\lket{\bigcdot}$ denotes a Liouville space vector. The vector $\newket{\boldsymbol{\rho}(t)}$ is then governed by the equation of motion
\begin{align}
\partial_t \newket{\boldsymbol{\rho}(t) }= \boldsymbol{\hat{\mathcal{L}}} \newket{\boldsymbol{\rho}(t)},
\end{align}
%
where $\boldsymbol{\hat{\mathcal{L}}}$ is the time-independent  Liouville superoperator that couples the elements of the auxiliary density operators in a time-independent fashion. Using the same arguments as in the main text, it is straightforward to show that $\newket{\boldsymbol{\delta \rho}^{(m)}} \equiv \int_{0}^{\infty} dt \ t^m (\newket{\boldsymbol{\rho}(t)}-\newket{\boldsymbol{\rho}_s})$ satisfies
\begin{align}
\boldsymbol{\hat{\mathcal{L}}} \ket{\boldsymbol{\delta \rho}^{(m)}} = - \newket{\boldsymbol{c}^{(m)}} \equiv - \begin{cases}
\newket{\boldsymbol{\rho}_0 } - \newket{\boldsymbol{\rho}_s}, & \text{if } m= 0\\
m \cdot \ket{\boldsymbol{\delta \rho}^{(m-1)}},    &\text{if } m \neq 0.
\label{eq:Am0}
\end{cases}
\end{align}
The system operator $\delta \hat{\rho}^{(m)}$ is then given by $\delta \hat{\rho}^{(m)}_{\mathbf{n=0}}$. The trace of $\delta \hat{\rho}^{(m)}_{\mathbf{n=0}}$ must be zero, and so the superoperator $w\boldsymbol{\hat{\mathcal{T}}}$ may be added to the left-hand side, as discussed above. In this case the effect of $\boldsymbol{\hat{\mathcal{T}}}$ is given by $\boldsymbol{\hat{\mathcal{T}}} \newket{\boldsymbol{v}} \propto \mathrm{Tr}[\hat{v}_{\mathbf{n}=\mathbf{0}}]$ for any $\newket{\boldsymbol{v}}$. Assuming that $\mathrm{Tr} [ (\boldsymbol{\hat{\mathcal{L}}} \ket{\boldsymbol{v}} )_{\mathbf{n} =\mathbf{0} } ] = 0$ for any $\newket{\boldsymbol{v}}$ and applying the same argument described in Appendix \ref{app:non_markov} shows that adding $w\boldsymbol{\hat{\mathcal{T}}}$ implies both Eq. (\ref{eq:Am0}) and $\mathrm{Tr}[\delta \hat{\rho}^{(m)}_{\mathbf{n} =\mathbf{0}  }] = 0$, as required. The condition $\mathrm{Tr} [ (\boldsymbol{\hat{\mathcal{L}}} \ket{\boldsymbol{v}} )_{\mathbf{n} =\mathbf{0} } ] = 0$ can be proved directly from the HEOM, which are given by
 \cite{heom_strumpfer}
\begin{align}
& \partial_t \hat{\rho}_{\mathbf{n}} = -\frac{i}{\hbar}[\hat{H}_{\mathrm{S}}, \hat{\rho}_{\mathbf{n}} ] - \sum_{a=1}^{M} \sum_{k=0}^{K} n_{ak} \nu_{ak} \hat{\rho}_{\mathbf{n}} - \sum_{a=1}^{M} \left( \frac{2 \lambda_a }{\hbar^2 \beta \gamma_a} -\sum_{k=0}^{K} \frac{c_{ak} }{\hbar \nu_{ak} }\right) [ \hat{F}_a, [\hat{F}_a,\hat{\rho}_{\mathbf{n}} ]] \nonumber \\
& -i \sum_{a=1}^{M}[ \hat{F}_a, \sum_{k=0}^{K} \hat{\rho}_{\mathbf{n}^+_{ak}} ] - \frac{i}{\hbar} \sum_{a=1}^{M} \sum_{k=0}^{K} n_{ak}(c_{ak} \hat{F}_a \hat{\rho}_{\mathbf{n}_{ak}^{-}} -c_{ak}^{*}\hat{\rho}_{\mathbf{n}_{ak}^{-}} \hat{F}_a ). \label{eq:heom}
\end{align}
Here the vector $\mathbf{n}$ that characterizes the hierarchy of auxiliary density matrices is defined by its elements $\{ \mathbf{n}_{ak} \}$, with $a \ \in [1,M]$ ($M$ is the number of baths) and $k \in [0, K]$, with $K$ the maximum number of Matsubara terms used in the bath correlation function. The vector $\mathbf{n}_{ak}^{\pm}$ is shorthand for $ (n_{10}, ... , n_{ak} \pm 1, ... , n_{MK} )$, $\nu_{ak}$ are the Matsubara frequencies, and $\beta$ is the inverse temperature. Bath $a$ couples to the system operator $\hat{F}_a$ through a Drude spectral density, with the reorganization energy $\lambda_a$ and inverse correlation time $\gamma_a$. Applying the cyclic property of the trace and using the fact that the second term in Eq. (\ref{eq:heom}) is zero for $\mathbf{n} =\mathbf{0}$ shows that indeed $\mathrm{Tr} [ (\boldsymbol{\hat{\mathcal{L}}} \ket{\boldsymbol{v}} )_{\mathbf{n} =\mathbf{0} } ] = 0$. Note that the hierarchy may be terminated using the so-called time non-local truncation, wherein all auxiliary matrices of a certain hierarchy are set to zero \cite{better_markov_heom1}, or with the time-local truncation that applies a Markovian approximation to these matrices \cite{heom_time_local,heom_strumpfer}.
In order to calculate progress moments under the dynamics induced by the HEOM, we assume that a truncation scheme has been chosen that ensures that $\boldsymbol{\hat{\mathcal{L}}}$ is time-independent, such as the time non-local scheme.
\subsection{Alternative formulation based on Laplace transforms}
\label{app:laplace}
\setcounter{equation}{0}
Here we discuss an alternative method of reconstructing the dynamics based on Laplace transforms. We have shown in the main text that progress moments can be used to reconstruct the dynamics by projecting the progress variable onto an exponential basis. The decay rates and expansion coefficients of the projection are then obtained numerically from the progress moments. An alternative approach is to consider the Laplace transform of the progress variable:
\begin{align}
\int_{0}^{\infty} dt \ \braket{\hat{\chi}(t)} e^{-k_n t} = \sum_m f_m /(k_n + k_m), \label{eq:exp_basis_classic}
\end{align}
where we have expressed the progress variable in the exponential basis of Eq. (\ref{eq:expansion}). If the $k_n$ and $k_m$ are chosen \textit{a priori}, and if the left-hand side is known, then Eq. (\ref{eq:exp_basis_classic}) can be solved for the $f_m$. The left-hand side can be evaluated for the time convolution master equation $\partial_t \hat{\rho}(t) = \int_{0}^{t} d\tau \ \hat{\mathcal{K}}(t-\tau) \hat{\rho}(\tau) $, yielding \cite{jesenko} 
\begin{align}
\int_{0}^{\infty} dt \ \hat{\rho}(t) e^{-k_n t} =  [k_n \hat{\mathds{1}} - \hat{\mathcal{L}}(k_n) ]^{-1} \hat{\rho}_0, \label{eq:laplace}
\end{align}
where $\hat{\mathcal{L}}(k_n) = \int_{0}^{\infty} dt \  e^{-k_nt} \hat{\mathcal{K}}(t)$ is the Laplace transform of $\hat{\mathcal{K}}(t)$ at the rate $k_n$. The analogous equation to Eq. (\ref{eq:laplace}) in the HEOM formalism is
\begin{align}
\int_{0}^{\infty} dt \ \newket{\boldsymbol{{\rho}}(t)}  e^{-k_n t} =  [k_n \boldsymbol{\hat{\mathds{1}}} - \boldsymbol{\hat{\mathcal{L}}} ]^{-1} \newket{\boldsymbol{{\rho}}_0}. \label{eq:laplace2}
\end{align}
The time-dependent memory kernel may be obtained from propagation of the HEOM, as discussed above, while the Markovian contribution $\int_{0}^{\infty} dt \ \hat{\mathcal{K}}(t)$ can be calculated from a set of eigenvectors associated with $\boldsymbol{\hat{\mathcal{L}}}$ \cite{jesenko2}.

The benefit of this method is that the Laplace transform of the progress variable at each rate $k_n$ is independent of all other Laplace transform at rates $k_n \neq k_m$. This is in contrast with the progress moments, which must be calculated sequentially. Hence the calculation of the Laplace transforms is amenable to parallel computation, which can greatly enhance computational efficiency. The independence of the transformations also ensures that there is no sequential build-up of numerical error. 
The drawback is that the $k_n$ must be chosen before calculating the Laplace transforms. This naturally necessitates the use of more basis functions than if the $k_n$ are calculated from progress moments. The power of the latter technique is that the relevant decay rates emerge directly from the progress moments, so that only a few exponential functions must be used. In the Laplace transform technique the relevant decay rates are not known, and so the function of interest must be approximated by larger sums of weighted exponentials with arbitrarily chosen decay rates.  

As an aside, consider that if the dynamics of interest occur on a time scale $t_{\mathrm{chem}} \gg \tau_{\mathrm{B}}$, where $\tau_{\mathrm{B}}$ is the bath relaxation time, only rates satisfying $k_n^{-1} \ \mathcal{O}(t_{\mathrm{chem}}) \gg \tau_{\mathrm{B}}$ will be needed to reconstruct the gross dynamics of the progress variable. Since the kernel $\hat{\mathcal{K}}(t)$ decays on a time scale $\tau_{\mathrm{B}}$, in practice $\hat{\mathcal{L}}(k_n)$ can be replaced with $\int_{0}^{\infty} dt \ \hat{\mathcal{K}}(t) = \hat{\mathcal{L}}_0$, which is simply the Markovian contribution to the dynamics. This justifies the focus on Markovian dynamics in the main text, which is relevant for systems that evolve on the slow time scale $t_{\mathrm{chem}}$. Indeed this is, in some sense, a generalization of the normal Markovian approximation, since it compares $\tau_{\mathrm{B}}$ to the rate of the process, $t_{\mathrm{chem}}$, as opposed to comparing $\tau_{\mathrm{B}}$ to all rates in the system.

A hybrid of the Laplace transform technique and the progress variable technique provides an efficient way of reconstructing the dynamics characterized by two gross time scales. To see this, consider the case that a single rate $k_0$ characterizes the fast dynamics of $\braket{\hat{O}(t)}$, and that all other decay rates $k_{n\neq 0}$ are far slower:
\begin{align}
\braket{\hat{O}(t)} = f_0 e^{-k_0 t} + \sum_{n \neq 0} f_n e^{-k_nt} + \mathrm{Tr}[\hat{\rho}_s \hat{O}],  \ \ \ \ k_0 \gg k_{n \neq 0}.
\end{align}
Then for $k_0^{-1} \ll t \ll k_{n \neq 0}^{-1}$, the variable $\braket{\hat{O}(t)}$ reaches a quasi-stationary value, $O_{\mathrm{qs}} = \mathrm{Tr}[\hat{\rho}_s \hat{O}] + \sum_{n\neq 0} f_n $. Since the progress variable technique most accurately reconstructs late time dynamics, we assume that the $f_n$ and $k_n$ are accurate for $n \neq 0$, but that the accuracy of $k_0$ requires improvement. Denoting the true fast rate as $\tilde{k}_0$ and applying a Laplace transform at the rate $k_0$ then yield
\begin{align}
\int_{0}^{\infty} dt \ e^{-k_0 t} (\braket{\hat{O}} - O_{\mathrm{qs}}) = \mathrm{Tr}[ ([k_0 \hat{\mathds{1} }- \hat{\mathcal{L}}]^{-1} \hat{\rho}_0 )\hat{O}] - k_0^{-1} O_{\mathrm{qs}} = \frac{f_0}{\tilde{k}_0+k_0},
\end{align}
which can be solved for $\tilde{k}_0$ by using $O_{\mathrm{qs}} = \mathrm{Tr}[\hat{\rho}_s \hat{O}] + \sum_{n\neq 0} f_0$.
\microtypesetup{protrusion=false}
\bibliography{thesis_bib}

\begin{thebibliography}{72}
\bibitem[{van()}]{vanishing_footnote}
\bibinfo{note}{We are implicitly assuming that $\hat{\rho}(t) \to \hat{\rho}_s$
  faster than $t^m \to \infty$. This must be the case if the progress moment
  exists.}
\bibitem[{T_f()}]{T_footnote}
\bibinfo{note}{To visualize $\hat{\mathcal{T}}$, consider deforming $\delta
  \hat{\rho}_n$ into a vector $\lket{\delta \rho_n}$, such that $\lket{\delta
  \rho_n} = [ (\delta \hat{\rho}_n)_{11}, \ (\delta \hat{\rho}_n)_{12}, \
  (\delta \hat{\rho}_n)_{13}, ..., (\delta \hat{\rho}_n)_{1d}, (\delta
  \hat{\rho}_n)_{21}, (\delta \hat{\rho}_n)_{22}, ... ]^{T}$, where $ (\delta
  \hat{\rho}_n)_{ij} = \braket{i \vert \delta \hat{ \rho}_n \vert j}$, $d$ is
  the dimension of the Hilbert space, and $T$ denotes matrix transposition.
  Then $\hat{\mathcal{T}}$ is a matrix with components $\mathcal{T}_{\alpha
  \beta}$, where $\alpha = (i-1) d + j $ and $\beta = (k-1)d + l$. Since
  $\hat{\mathcal{T}}$ is non-zero only for $i=j = 1$, $\mathcal{T}_{\alpha
  \beta}$ is zero except when $\alpha=1$. Moreover, since $\hat{\mathcal{T}}$
  is non-zero only for $k=l$, we see that $\hat{\mathcal{T}}$ is a matrix with
  ones in its upper row, at locations corresponding to the elements $(\delta
  \hat{\rho}_n)_{11}$, $(\delta \hat{\rho}_n)_{22}$, and so on. Hence
  $\hat{\mathcal{T}} \lket{\delta \rho_n} = \{\sum_i (\delta
  \hat{\rho}_n)_{ii}\} \times [ 1, 0 , 0, ... , 0]^{T} = \mathrm{Tr}(\delta
  \hat{\rho}_n) \times [ 1, 0 , 0, ... , 0]^{T} $. Reforming $\hat{\mathcal{T}}
  \lket{\delta \rho_n}$ into a Hilbert space operator yields $\hat{\mathcal{T}}
  [\delta \hat{\rho}_n] = \mathrm{Tr}[\delta \hat{\rho}_n] \ket{1} \bra{1}$. If
  only the diagonal elements of $\delta \hat{\rho}_n$ are of interest, such as
  in the secular approximation, then $\lket{\delta \rho_n}$ can be written as
  $\lket{\delta \rho_n} = [(\delta \hat{\rho}_n)_{11} , (\delta
  \hat{\rho}_n)_{22}, ... , (\delta \hat{\rho}_n)_{dd} ]^{T}$. In this case,
  $\mathcal{T}_{\alpha \beta} = \delta_{\alpha 1}$, i.e. $\hat{\mathcal{T}}$ is
  a matrix with ones at each location in the upper row.}
\bibitem[{ns_()}]{ns_footnote}
\bibinfo{note}{To understand this notation, consider an example in which
  $\hat{\kappa}$ is defined relative to a property of $\hat{\mathcal{L}}$, such
  as the steady state $\hat{\rho}_s$, which is the solution to the equation
  $\hat{\mathcal{L}} \hat{\rho}_s = \hat{0}$. An example would be $\hat{\kappa}
  \hat{X} = \mathrm{Tr}[\hat{\rho}_s \hat{X}] \hat{\rho}_s$ for any operator
  $\hat{X}$. The effect of $\hat{\kappa}\hat{X}$ in the equation $(
  \hat{\mathcal{L}} + \hat{\kappa} ) \hat{X} = \hat{v}$ would then be to
  enforce the constraint $\mathrm{Tr}[\hat{\rho}_s \hat{X}] =
  \mathrm{Tr}[\hat{v}] / \mathrm{Tr}[\hat{\rho}_s]$. This follows from taking
  the trace of both sides of the equation and noting that
  $\mathrm{Tr}[\hat{\mathcal{L}} \hat{X} ] = 0$ for any operator $\hat{X}$
  (proved in Appendix \ref{app:non_markov}). Similarly, $\hat{\kappa}_{1-\eta}$
  would be defined with respect to the steady state $(\hat{\rho}_s)_{1-\eta}$
  obtained from solving $[\hat{\mathcal{L}}_{\mathrm{sec}} +
  (1-\eta)\hat{\mathcal{L}}_{\mathrm{ns}} ] (\hat{\rho}_s)_{1-\eta} = \hat{0}$.
  In the cases studied here, $\hat{\kappa}$ is not defined relative to any
  specific property of $\hat{\mathcal{L}}$, and so $\hat{\kappa} =
  \hat{\kappa}_{1-\eta}$. However, we include $\hat{\kappa}_{1-\eta}$ to keep
  the analysis general, for the following reason. As discussed above, the
  presence of $\hat{\kappa}$ renders the sum $(\hat{\mathcal{L}} +
  \hat{\kappa})$ non-singular. If $\hat{\kappa}$ is defined with respect to a
  specific property of $\hat{\mathcal{L}}$, such as the steady state
  $\hat{\rho}_s$, then the sum $(\hat{\mathcal{L}}_{1-\eta} + \hat{\kappa})$
  may not be non-singular in general, and so $\hat{G}^{-1}$ may not exist. By
  contrast, the sum $(\hat{\mathcal{L}}_{1-\eta} + \hat{\kappa}_{1-\eta})$ is
  non-singular by construction, and so the inverse $\hat{G}^{-1}$ exists.}
\bibitem[{dec()}]{decay_vs_growth}
\bibinfo{note}{For example, if the progress variable decays exponentially, such
  that $\braket{\hat{\chi}(t)} = e^{-kt}$, then the $n$th progress moment is
  $I_n = n! \hspace*{0.05cm} k^{-(1+n)}$. If $k^{-1} = 2 \ \mathrm{ms}$, and
  one uses units of microseconds, then $I_{n} \to \infty$ as $n \to \infty$. If
  one uses units of seconds, then $I_{n} \to 0$ as $n \to \infty$.}
\bibitem[{foo()}]{footnote}
\bibinfo{note}{If one requires more basis functions than $\mathcal{M}$, seed
  values for $m +1>\mathcal{M}$ can be obtained by substituting the
  $m_{\mathrm{max}} + 1=\mathcal{M}$ solution into the remaining equations, and
  solving for $f_{\mathcal{M}}$ and $k_{\mathcal{M}}$ approximately.}
\end{thebibliography}
\end{document}